\documentclass{article}

\begin{document}

\title{Acceleration-dependent self-interaction effects as a basis for inertia}
\author{Vesselin Petkov \\
Science College, Concordia University\\
1455 de Maisonneuve Boulevard West\\
Montreal, Quebec H3G 1M8\\
vpetkov@alcor.concordia.ca}
\date{}
\maketitle

\begin{abstract}
The paper pursues two aims. First, to revisit the classical
electromagnetic mass theory and develop it further by making use
of a corollary of general relativity - that the propagation of
light in non-inertial reference frames is anisotropic. Second, to
show that the same type of acceleration-dependent self-interaction
effects that give rise to the inertia and mass of the classical
electron appear in quantum field theory as well when the general
relativistic frequency shift of the virtual quanta, mediating the
electromagnetic, weak, and strong interactions between
non-inertial particles, is taken into account. Those effects may
account for the origin of inertia and mass of macroscopic objects.
\end{abstract}

\section{Introduction}

Recently there has been a renewed interest in the nature of inertia \cite%
{haisch}-\cite{woodward}. This is not surprising since the issue of inertia
along with that of gravitation have been the most outstanding puzzles in
physics for centuries. Even now, at the beginning of the twenty first
century, the situation is the same - the nature of inertia remains an
unsolved mystery in modern physics; our understanding of gravity can be
described in the same way since the modern theory of gravitation, general
relativity, added very little to our understanding of the mechanism of
gravitational interaction. The mystery of gravity has been even further
highlighted by the fact that general relativity, which provides a consistent
no-force explanation of gravitational interaction of bodies following
geodesic paths, is silent on the nature of the very force we identify as
gravitational - the force acting upon a body deviated from its geodesic path
while being at rest in a gravitational filed.

In the past there have been two major and very different attempts to
understand what causes inertia. In 1881 Thomson \cite{thomson} first
realized that a charged particle was more resistant to being accelerated
than an otherwise identical neutral particle and conjectured that inertia
can be reduced to electromagnetism. Owing mostly to the works of Heaviside
\cite{heaviside}, Searle \cite{searle}, Abraham \cite{abraham}, Lorentz \cite%
{lorentz}, Poincar\'{e} \cite{poincare}, Fermi \cite{fermi21, fermi22},
Mandel \cite{mandel}, Wilson \cite{wilson}, Pryce \cite{pryce}, Kwal \cite%
{kwal} and Rohrlich \cite{rohrlich, rohrlich90} this conjecture was
developed in the framework of the classical electron theory into what is now
known as the classical electromagnetic mass theory of the electron. In this
theory inertia is regarded as a local phenomenon originating from the
interaction of the electron with its own electromagnetic field \cite{emmnote}%
. Around 1883 Mach \cite{mach} argued that inertia was caused by all the
matter in the Universe thus assuming that the local property of inertia had
a non-local cause.

While a careful theoretical analysis \cite{weinberg} speaks against Mach's
hypothesis, the electromagnetic mass approach to inertia, on the contrary,
is still the only theory that predicts the experimental fact that at least
part of the inertia and inertial mass of every charged particle is
electromagnetic in origin. As Feynman put it: \textquotedblright There is
definite experimental evidence of the existence of electromagnetic inertia -
there is evidence that some of the mass of charged particles is
electromagnetic in origin\textquotedblright\ \cite{feynman}. And despite
that at the beginning of the twentieth century many physicists recognized
\textquotedblright the tremendous importance, which the concept of
electromagnetic mass possesses for all of physics\textquotedblright\ since
\textquotedblright it is the basis of the electromagnetic theory of
matter\textquotedblright\ \cite{fermi22a} it has been inexplicably abandoned
after the advent of relativity and quantum mechanics. And that happened even
though the classical electron theory predicted \emph{before} the theory of
relativity that the electromagnetic mass increases with the increase of
velocity, yielding the correct velocity dependence, and that the
relationship between energy and mass is $E=mc^{2}$ \cite[pp. 28-3, 28-4]%
{feynman}, \cite{4/3factor}. Now \textquotedblright the state of the
classical electron theory reminds one of a house under construction that was
abandoned by its workmen upon receiving news of an approaching plague. The
plague in this case, of course, was quantum theory. As a result, classical
electron theory stands with many interesting unsolved or partially solved
problems\textquotedblright\ \cite{pearle}.

The purpose of this paper is to demonstrate that the classical electron
theory (more specifically, the classical electromagnetic mass theory)
considered in conjunction with general relativity sheds some light on the
nature of inertia and mass.

Inertia is the resistance a particle offers to its acceleration and inertial
mass is defined as the measure of that resistance. The classical
electromagnetic mass theory shows that an accelerated charge resists the
deformation of its electromagnetic field (caused by the accelerated motion)
and the measure of this resistance is the inertial electromagnetic mass of
the charge $m^{a}=U/c^{2}$, where $U$ is the energy of the electromagnetic
field of the charge. However, when the field is considered in the
accelerated reference frame $N^{a}$ in which the charge is at rest it is not
clear how the acceleration causes the distortion of the charge's field in $%
N^{a}$. The same difficulty is encountered when one calculates the electric
field in the non-inertial reference frame $N^{g}$ of a charge supported in a
gravitational field. Both problems are resolved if a corollary of general
relativity - that the propagation of electromagnetic signals (for short
light) in non-inertial frames of reference is anisotropic - is taken into
account. As we shall see in Section 4 the average velocity of light between
two points in a non-inertial frame ($N^{a}$ or $N^{g}$) is anisotropic - it
depends on from which point it is determined (since at that point the local
speed of light is always $c$). When two observers in $N^{a}$ and $N^{g}$
calculate the electric fields of two charges at rest in $N^{a}$ and $N^{g}$,
respectively, they find that the fields are as distorted as required by the
equivalence principle (see Section 5). The charges resist the deformation of
their fields through the self-forces $\mathbf{F}_{self}^{a}=-m^{a}\mathbf{a}$
and $\mathbf{F}_{self}^{g}=m^{g}\mathbf{g}$, where $m^{a}=U/c^{2}$ and $%
m^{g}=U/c^{2}$, again in agreement with the equivalence principle.

As will be shown in Section 5 the only way for a charge in $N^{a}$ or $N^{g}$
to prevent its field from getting distorted is to compensate the anisotropy
in the propagation of light there by falling with an acceleration $\mathbf{a}
$ or $\mathbf{g}$, respectively. If the charge is prevented from falling,
its field distorts which gives rise to the self-force $\mathbf{F}_{self}^{a}$
or $\mathbf{F}_{self}^{g}$. This self-force, in turn, acts back on the
charge and tries to force it to fall in order to compensate the anisotropy
in the propagation of light and to restore the Coulomb shape of its field.
This shows that the classical electromagnetic mass theory reveals a close
connection between the shape of the electric field of a charge and its state
of motion. If the field of a charge is the Coulomb field (i.e. it is not
distorted), the charge offers no resistance - it moves in a non-resistant
way (by inertia). However, if the charge accelerates, its field deforms and
its motion with respect to an inertial reference frame $I$ is resistant;
when observed in $N^{a}$ the charge resists the deformation of its field and
its being prevented from falling in $N^{a}$. If the field of a charge is
distorted due to its being at rest in a gravitational filed (i.e. at rest in
$N^{g}$), the charge also resists the deformation of its field and its being
prevented from falling in the gravitational field. In other words, if the
field of a charge is the Coulomb field, the charge is represented by a
geodesic worldline; if the field is distorted, the worldline of the charge
is not geodesic.

The classical electromagnetic mass theory presents an intriguing account of
the origin of inertia and mass (and the equivalence of the inertial mass $%
m^{a}$ and the passive gravitational mass $m^{g}$) of a charge in terms of
its self-interaction with its distorted electromagnetic field. One problem
with the classical theory, since it does not take into account the strong
and weak interactions, is its prediction that the \textit{entire} mass of a
charge should be electromagnetic in origin. However, if the electromagnetic
interaction gives rise to inertia and electromagnetic mass, the strong and
weak interactions as fundamental forces should make a contribution to the
mass as well \cite{stephani}.

The study of the classical electromagnetic mass theory makes it possible to
ask whether the same type of acceleration-dependent self-interaction effects
in quantum field theory (QFT) give rise to inertia and mass. All
interactions in QFT are realized through the exchange of virtual quanta
constituting the corresponding "fields". Consider again the electromagnetic
interaction in quantum electrodynamics (QED). In QED the quantized electric
field of a charge is represented by a cloud of virtual photons that are
constantly being emitted and absorbed by the charge. It is believed that the
attraction and repulsion electric forces between two charges interacting
through exchange of virtual photons originate from the recoils the charges
suffer when the virtual photons are emitted and absorbed.

A free (inertial) charge in QED is subjected to the recoils resulting from
the emitted and absorbed virtual photons which constitute its own electric
field. Due to spherical symmetry, all recoils caused by both the emitted and
absorbed virtual photons cancel out exactly and the charge is not subjected
to any self-force. Hence, in terms of QED a charge is represented by a
geodesic worldline if the recoils from the emitted and absorbed virtual
photons completely cancel out.

As it is the momentum of a photon that determines the recoil felt by a
charge when the photon is emitted or absorbed, the recoils resulting from
the virtual photons emitted by a non-inertial charge also cancel out since,
as seen by the charge, all photons are emitted with the same frequencies
(and energies) and therefore the same momenta. However, the frequencies of
the virtual photons coming from different directions before being absorbed
by a non-inertial charge are direction dependent (blue or red shifted). It
has not been noticed so far that this directional dependence of the
frequencies of the virtual photons absorbed by a non-inertial charge
disturbs the balance of the recoils to which the charge is subjected. In
turn, that imbalance gives rise to a self-force which acts on the
non-inertial charge. It should be specifically stressed that the mechanism
which gives rise to that self-force is \textit{not} hypothetical - it is the
accepted mechanism responsible for the origin of attraction and repulsion
forces in QED.

The self-force, resulting from the imbalance in the recoils caused by the
virtual photons absorbed by a non-inertial charge, is a resistance force
since it acts only on non-inertial charges. It arises only when an inertial
charge is prevented from following a geodesic worldline; no self-force is
acting on an inertial charge (following a geodesic worldline).

As we shall see in Section 6 in the case of an accelerating charge the
resistance self-force has the form of the inertial force which resists the
deviation of the charge from its geodesic path. When a charge is supported
in a gravitational field the self-force also resists the deviation of the
charge from its geodesic path and has the form of what is traditionally
called the gravitational force.

It is clear that non-inertial weak and strong (color) charges will be also
subjected to a self-force arising from the imbalance in the recoils caused
by the absorbed $W$ and $Z$ particles in the case of weak interaction and
the absorbed gluons in the case of strong interaction. Therefore it appears
that the acceleration-dependent self-interaction effects in QFT are similar
to those in the classical electromagnetic theory and may account for the
origin of inertia and mass.

The picture which emerges is the following. Consider a body whose
constituents are subjected to electromagnetic, weak and strong interactions.
If the recoils from all virtual quanta (photons, $W$ and $Z$ particles, and
gluons) mediating the interactions cancel out precisely, the body is
represented by a geodesic worldline; it is offering no resistance to its
motion and is therefore moving by inertia. When the body is accelerating the
balance of the recoils caused by the absorbed virtual quanta is disturbed
which gives rise to a self-force. The worldline of a body whose constituents
have distorted electromagnetic, weak, and strong fields is not geodesic (the
distortion of the "fields" manifests itself in the fact that the recoils
from the absorbed virtual quanta do not cancel out). As a result the body
resists the deformation of its electromagnetic, weak, and strong "fields"
and therefore its acceleration which is causing the deformation. The
self-force is a resistance force and is composed of three components -
electromagnetic, weak and strong. Therefore both inertia and inertial mass
appear to originate from the lack of cancellation of the recoils caused by
the absorbed virtual quanta mediating the electromagnetic, weak, and strong
interactions.

If the body is at rest in a gravitational field, the frequencies (i.e. the
energies) of the virtual quanta being absorbed by its constituent particles
are shifted. As a result the recoils from the virtual photons, $W$ and $Z$
particles, and gluons which every constituent particle of the body suffers
do no cancel out. That imbalance in the recoils gives rise to a self-force
which has the form of the gravitational force (as shown in Section 6) and is
also composed of three components - electromagnetic, weak and strong. This
means that the passive gravitational mass, like the inertial mass, appears
to originate from the imbalance in the recoils caused by the absorbed
virtual quanta mediating the electromagnetic, weak, and strong interactions.
Therefore this picture provides a natural explanation of the equivalence of
inertial and passive gravitational masses - they have the \textit{same}
origin. The anisotropy in the propagation of the virtual quanta is
compensated if the body falls with an acceleration $\mathbf{g}$. The recoils
from all absorbed virtual quanta (photons, $W$ and $Z$ particles, and
gluons) the falling body suffers cancel out exactly and the body moves in a
non-resistant way (following a geodesic path). This mechanism offers a nice
explanation of why all bodies fall in a gravitational field with the \textit{%
same} acceleration.

The outlined picture suggests that inertia and the entire inertial and
passive gravitational mass originate from acceleration-dependent
self-interaction effects in QFT - the constituent particles of every
non-inertial body are subjected to a self-force which is caused by the
imbalance in the recoils from the absorbed virtual quanta. What spoils the
picture is the rest mass of the $Z$ particle. The unbalanced recoils from
this particle explains the contribution of the weak interaction to the mass
of every particle undergoing weak interactions in which the $Z$ particle is
involved. However, what accounts for the mass of the $Z$ particle which is
one of the carriers of this interaction remains a mystery.

On the one hand, it follows from QFT, when the general relativistic
frequency shift is taken into account, that electromagnetic, weak, and
strong interactions all make contributions to inertia and mass. On the other
hand, the fact that the $Z$ particle involved in mediating the weak
interaction possesses a rest mass demonstrates that not all mass is composed
of electromagnetic, weak, and strong contributions. Obviously, it will be
the experiment that will determine how much of the mass is due to
electromagnetic, weak, and strong interactions, and how much is caused by
the Higgs or another unknown mechanism.

One obvious question that has remained unanswered so far is about the
gravitational interaction. If we manage to quantize gravitation and the
existence of gravitons is confirmed, gravitational interaction will make a
contribution to the mass as well, and more importantly may account for the
mass of the $Z$ particle.

The paper addresses two main questions: (i) Are inertia and both inertial
and gravitational mass of the classical electron fully explained by the
electromagnetic mass theory? and (ii) Do the electromagnetic, weak, and
strong interactions, in the framework of QFT, all contribute to inertia and
mass?

Section 2 discusses the reasons why this paper starts with the study of the
classical electron. Section 3 examines the arguments against the classical
electromagnetic mass theory. Section 4 deals with an important but
overlooked up to now corollary of general relativity - that in addition to
the coordinate velocity of light one needs two different average velocities
(coordinate and proper) to account fully for the propagation of light in
non-inertial reference frames. In Section 5 it is shown that (i) the inertia
and mass of the electron are caused by acceleration-dependent
electromagnetic self-interaction effects, and (ii) the inertial and
gravitational mass of the classical electron are purely electromagnetic in
origin (which naturally explains their equivalence). Section 6 applies the
mechanism of exchange of virtual quanta that gives rise to attraction and
repulsion forces in QFT to the case of a non-inertial charge.

\section{Why the classical electron?}

As mentioned in the Introduction the study of the inertial properties of the
classical electron reveals that its inertia and mass originate from the
interaction of the electron charge with its own distorted field. In Section
5 we will demonstrate that this is really the case and then in Section 6
will show that in QFT the \textit{same} mechanism - interaction of electric,
weak, and strong charges with their distorted "fields" - gives rise to
contributions from the electromagnetic, weak, and strong interactions to
inertia and mass of all bodies.

Although as it will become clear throughout the paper that it is the
analysis of the inertial properties of the classical electron that provided
the hint of how to approach the issue of inertia in QFT, let me briefly
explain why this paper starts with the study of inertia and mass of the
classical electron.

Often the first reaction to any study of the classical model of the electron
(a small charged spherical shell) questions why it should be studied at all
since it is clear that this model is wrong: the classical electron radius
that gives the correct electron mass is $\sim 10^{-15}m$ whereas experiments
probing the scattering properties of the electron found that its size is
smaller than $10^{-18}m$ \cite{bender}. Unfortunately, it is not that
simple. An analysis of why the electron does not appear to be so small has
been carried out by Mac Gregor \cite{macgregor}. However, what immediately
shows that the scattering experiments do not tell the whole story is the
fact that they are relevant only to the particle aspect of the electron.

Despite all studies specifically devoted to the nature of the electron (see,
for instance, \cite{electron91}, \cite{electron97}) no one knows what an
electron looks like before being detected and some even deny the very
correctness of such a question. One thing, however, is completely clear: the
experimental upper limit of the size of the electron ($<10^{-18}m$) cannot
be interpreted to mean that the electron is a particle (localized in a
region whose size is smaller than $10^{-18}m$) without contradicting both
quantum mechanics and the existing experimental evidence. Therefore, the
scattering experiments tell very little about what the electron itself is
and need further studies in order to understand their meaning. For this
reason those experiments are not an argument against any study of the
classical model of the electron.

As one of the most difficult problems of the classical electron is its
stability one may conclude that the basic assumption in the classical model
of the electron - that there is \textit{interaction} between the elements of
its charge through their distorted fields - may be wrong. The very existence
of the radiation reaction force, however, seems to imply that there is
indeed interaction (repulsion) between the different "parts" of the electron
charge. The radiation reaction is due to the force of a charge on itself -
the net force exerted by the fields generated by different parts of the
charge distribution acting on one another \cite[p. 439]{griffiths}. In the
case of a \emph{single} radiating electron the presence of a radiation
reaction force appears to suggest that there is interaction of different
"parts" of the electron.

Here are two more reasons justifying the analysis of the classical electron:

(i) The calculations of the inertial and gravitational forces acting on a
non-inertial classical electron (accelerating and at rest in a gravitational
field, respectively) yield the correct expressions for these forces. This
means that Newton's second law can be derived on the basis of Maxwell's
equations and the classical model of the electron (see Sections 5.1 and
5.2). It is unlikely that such a result may be just a coincidence.

(ii) The completion of the classical electromagnetic mass theory by taking
into account the average anisotropic velocity of light in non-inertial
frames of reference is of importance for the following reason as well. As
inertia and gravitation have predominantly macroscopic manifestations it
appears natural to expect that these phenomena should possess not only a
quantum but a classical description as well. This expectation is
corroborated by the very existence of classical theories of gravitation -
Newton's gravitational theory and general relativity. In addition to
predicting the experimental evidence of the existence of electromagnetic
inertia and mass, the classical electromagnetic mass theory yields, as we
shall see, the correct expressions for the inertial and gravitational forces
acting on a non-inertial classical electron. Therefore its completion will
naturally make it the classical theory of inertia. It is worth exploring the
classical electromagnetic mass theory further since the results obtained, as
we shall see, may serve as guiding principles for our understanding of the
nature of inertia and mass in QFT. The most general guiding principle is
given by Bohr's correspondence principle which states that the quantum
theory must agree with the classical theory where the classical theory's
predictions are accurate. As the classical theory of inertia accurately
predicts the existence of electromagnetic inertia and mass of charged
classical particles the application of Bohr's correspondence principle
implies that the chances of any modern theory of inertia can be evaluated by
seeing whether it can be considered a quantum generalization of the
classical electromagnetic mass theory.

\section{Classical electromagnetic mass theory and the arguments against it}

According to the classical electromagnetic mass theory it is the \emph{%
unbalanced} repulsion of the volume elements of the charge caused by the
distorted field of an accelerating electron that gives rise to the
electron's inertia and inertial mass. Since the electric field of an
inertial electron (represented by a straight worldline in flat spacetime) is
the Coulomb field the repulsion of its charge elements cancels out exactly
and there is no net force acting on the electron. If, however, the electron
is accelerated its field distorts, the balance in the repulsion of its
volume elements gets disturbed, and as a result it experiences a net
self-force $\mathbf{F}_{self}$ which resists its acceleration - it is this
resistance that the classical electromagnetic mass theory regards as the
electron's inertia. The self-force is opposing the external force that
accelerates the electron (i.e. its direction is opposite to the electron's
acceleration $\mathbf{a}$) and turns out to be proportional to $\mathbf{a}$:
$\mathbf{F}_{self}^{a}$ $=-m^{a}\mathbf{a}$, where the coefficient of
proportionality $m^{a}$ represents the inertial mass of the electron and is
equal to $U/c^{2}$, where $U$ is the energy of the electron field; therefore
the electron inertial mass is electromagnetic in origin.

The electromagnetic mass of the classical electron can be calculated by
three independent methods \cite{griffithsowen}: (i) energy-derived
electromagnetic mass $m_{U}=U/c^{2}$, where $U$ is the field energy of an
electron at rest (when the electron is moving with relativistic velocities $%
v $ then $m_{U}=U/\gamma c^{2}$, where $\gamma =\left( 1-v^{2}/c^{2}\right)
^{-1/2}$), (ii) momentum-derived electromagnetic mass $m_{p}=p/v$, where $p$
is the field momentum when the electron is moving at speed $v$ (for
relativistic velocities $m_{p}=p/\gamma v$), and (iii) self-force-derived
electromagnetic mass $m_{s}=F_{self}/a$, where $F_{self}$ is the self-force
acting on the electron when it has an acceleration $a$ (for relativistic
velocities $m_{s}=F_{self}/\gamma ^{3}a$).

There have been two arguments against regarding the entire mass of charged
particles as electromagnetic in classical (non-quantum) physics:

(i) There is a factor of $4/3$ which appears in the momentum-derived and the
self-force-derived electromagnetic mass - $m_{p}=\frac{4}{3}m_{U}$ and $%
m_{s}=\frac{4}{3}m_{U}$ (the energy-derived electromagnetic mass $m_{U}$
does not contain that factor). Obviously, the three types of electromagnetic
masses should be equal.

(ii) The inertia and mass of the classical electron originate from the
unbalanced mutual repulsion of its "parts" caused by the distorted electric
field of the electron. However, it is not clear what maintains the electron
stable since the classical model of the electron describes its charge as
uniformly distributed on a spherical shell, which means that its volume
elements tend to blow up since they repel one another.

Feynman considered the $4/3$ factor in the electromagnetic mass expression a
serious problem since it made the electromagnetic mass theory (yielding an
incorrect relation between energy and momentum due to the $4/3$ factor)
inconsistent with the special theory of relativity: ''It is therefore
impossible to get all the mass to be electromagnetic in the way we hoped. It
is not a legal theory if we have nothing but electrodynamics'' \cite[p. 28-4]%
{feynman}. It seems he was unaware that the $4/3$ factor which appears in
the momentum-derived electromagnetic mass had already been accounted for in
the works of Mandel \cite{mandel}, Wilson \cite{wilson}, Pryce \cite{pryce},
Kwal \cite{kwal}, and Rohrlich \cite{rohrlich} (each of them removed that
factor independently from one another). The self-force-derived
electromagnetic mass has been the most difficult to deal with, persistently
yielding the factor of $4/3$ \cite{griffithsowen}. By a covariant
application of Hamilton principle in 1921 Fermi \cite{fermi21} first
indirectly showed that there was no $4/3$ factor in the self-force acting on
a charge supported in a gravitational field. Here we shall see how the
factor of $4/3$ is accounted for in the case of an electron at rest in an
accelerating reference frame $N^{a}$ and in a frame $N^{g}$ at rest in a
gravitational field of strength $\mathbf{g}$, described in $N^{a}$ and $%
N^{g} $, respectively. After the $4/3$ factor has been removed the
electromagnetic mass theory of the classical electron becomes fully
consistent with relativity and the classical electron mass turns out to be
purely electromagnetic in origin.

Since its origin a century ago the electromagnetic mass theory has not been
able to explain why the electron is stable (what holds its charge together).
This failure has been seen as an explanation of the presence of the $4/3$
factor and has been used as evidence against regarding its entire mass as
electromagnetic. To account for the $4/3$ factor it had been assumed that
part of the electron mass (regarded as mechanical) originated from
non-electric forces (known as the Poincar\'{e} stresses \cite{poincare})
which hold the electron charge together. It was the inclusion of those
forces in the classical electron model and the resulting mechanical mass
that compensated the $4/3$ factor \emph{reducing} the momentum-derived
electromagnetic mass from $\left( 4/3\right) m_{U}$ to $m_{U}$. This
demonstrates that the \emph{attraction} non-electric Poincar\'{e} forces
make a \emph{negative} contribution to the entire electron mass. It turned
out, however, that the $4/3$ factor was a result of incorrect calculations
of the momentum-derived electromagnetic mass as shown by Mandel \cite{mandel}%
, Wilson \cite{wilson}, Pryce \cite{pryce}, Kwal \cite{kwal}, and Rohrlich
\cite{rohrlich}. As there remained nothing to be compensated (in terms of
mass), if there were some unknown attraction forces responsible for holding
the electron charge together, their \emph{negative} contribution (as
attraction forces) to the electron mass would result in reducing it from $%
m_{U}$ to $\left( 2/3\right) m_{U}$. Obviously, there are two options in
such a situation - either to seek what this time compensates the negative
contribution of the Poincar\'{e} stresses to the mass or to assume that the
hypothesis of their existence was not necessary in the first place
(especially after it turned out that the $4/3$ factor does not appear in the
correct calculation of the momentum-derived electromagnetic mass). A strong
argument supporting the latter option is the fact that if there existed a
real problem with the stability of the electron, the hypothesis of the
Poincar\'{e} stresses should be needed to balance the mutual repulsion of
the volume elements not only of an electron moving with constant velocity
(as in the case of the momentum-derived electromagnetic mass), but also of
an electron at rest in its rest frame. This, however, is not the case since
when the electron is at rest there is no $4/3$ factor problem in the \emph{%
rest}-energy-derived electromagnetic mass of the electron. If the electron
charge tended to blow up as a result of the mutual repulsion of its
\textquotedblright parts\textquotedblright , it should do so not only when
it is moving at constant velocity but when is at rest as well. Somehow this
obvious argument has been overlooked.

Another indication that the stability problem does not appear to be a real
problem is that it does not show up (through the $4/3$ factor) in the
correct calculations of the self-force either. As Fermi \cite{fermi21}
showed and as we shall see here the Poincar\'{e} stresses are not needed for
the derivation of the self-force-derived electromagnetic mass since the $4/3$
factor which was present in previous derivations of the self-force turned
out to be a result of not including in the calculations of an anisotropic
volume element which arises due to the anisotropic velocity of light in the
non-inertial reference frames where the self-force is calculated.

All this implies that there is no real problem with the stability of the
electron. We do not know why. What we do know, however, is that if there
were a stability problem it would inevitably show up in \emph{all}
calculations of the energy-derived, momentum-derived, and self-force-derived
electromagnetic mass which is not the case. Obviously, there should be an
answer to the question why calculations based on the indisputably wrong
classical model of the electron (i) correctly describe its inertial and
gravitational behaviour (including the equivalence of its inertial and
passive gravitational mass), and (ii) yield the correct expressions for the
inertial and gravitational force. My guess of what that answer might be is
that there is something in the classical model which leads to the correct
results. Most probably, the spherical distribution of the charge. However,
it is not necessary to assume that that distribution is a \textit{solid}
spherical shell existing at every single instant as a solid shell (like the
\textit{macroscopic} objects we are aware of) - only in that case its
"parts" will repulse one another and the sphere will tend to explode. It is
not unthinkable to picture an $elementary$ charge as being spherical but not
solid (with continuous distribution of the charge) \cite{elementary}.

The fact that the $4/3$ factor has been accounted for and the stability
problem does not appear to be a real problem (since it does not show up (i)
in the \emph{rest}-energy-derived electromagnetic mass of the electron, and
(ii) in the calculations of the self-force) indicates that, in the case of
the classical electron, the arguments against regarding its inertia and
inertial mass as entirely electromagnetic in origin are answered. We shall
see that not only is its inertial mass electromagnetic, but also both its
passive gravitational and active gravitational masses are purely
electromagnetic in origin. This in turn fully explains the behavior of an
electron in flat spacetime and in a gravitational field, thus providing
answers to the following fundamental questions: (i) Why does an electron
moving with uniform velocity in flat spacetime not resist its motion by
inertia (since Galileo inertia has been a postulate)? (ii) Why does an
accelerating electron in flat spacetime resist its acceleration? (iii) Why
does an electron falling toward the Earth's surface not resist its
acceleration? (iv) Why does an electron at rest on the Earth's surface
resist its being prevented from falling?

\section{Propagation of light in non-inertial reference frames}

For an inertial observer $I$ the electric field of an accelerating electron
is doubly distorted due to (i) the Lorentz contraction, and (ii) its
acceleration. As we are interested in the effect of the acceleration on the
shape of the electric field of an accelerating electron throughout the paper
we will be considering its \emph{instantaneous} electric field with respect
to $I$ in order to separate the deformation of the electric field due to the
Lorentz contraction from the distortion caused by the electron's
acceleration.

As the instantaneous electric field of an accelerating electron is distorted
with respect to $I$ it follows that there is a self-force acting on the
electron. For an observer in an accelerating frame of reference $N^{a}$ in
which the electron is at rest, however, it appears at first glance that the
electron field is not distorted since it is at rest in $N^{a}$ which would
mean that the electron is subjected to no self-force. If this were the case,
there would be a problem: the inertial and the non-inertial observers would
differ in their observation on whether or not the electron is subjected to a
force; as the existence of a force is an absolute (observer-independent)
fact all observers (both inertial and non-inertial) should recognize it.

The same problem arises if an electron is at rest in the non-inertial
reference frame $N^{g}$ of an observer supported in a gravitational field.
As the electron is at rest in $N^{g}$ it will appear that its field should
not be distorted in $N^{g}$. For an inertial observer $I$ (falling in the
gravitational field), however, the electron is accelerating with respect to $%
I$ and therefore its field is distorted. This problem disappears when a
corollary of general relativity that the propagation of light is anisotropic
in non-inertial reference frames is taken into account in the calculation of
the electron potential and field there. As we shall see below, due to the
average anisotropic velocity of light in $N^{a}$ and $N^{g}$, the scalar
potential $\varphi ^{a}$ and $\varphi ^{g}$ and the vector potential $%
\mathbf{A}^{a}$ and $\mathbf{A}^{g}$ are distorted and when the electric
field of an electron at rest in $N^{a}$ and $N^{g}$ is calculated by using
the scalar potential it turns out as distorted as the field seen by the
inertial observer $I$ instantaneously at rest with respect to $N^{a}$ and $%
N^{g}$.

We shall now determine the average anisotropic velocity of light in $N^{a}$
and $N^{g}$ in order to calculate the potential and the electric field of an
electron there which in turn will enable us to determine the self-forces $%
\mathbf{F}_{self}^{a}$ and $\mathbf{F}_{self}^{g}$ in $N^{a}$ and $N^{g}$,
respectively.

It has been overlooked that the average velocity of light determined in an
accelerating reference frame $N^{a}$ is anisotropic due to the accelerated
motion of $N^{a}$. The anisotropy of light in an accelerating frame is a
direct consequence of the fact that acceleration is \emph{absolute} in a
sense that there is an absolute, observer-independent, difference between a
geodesic worldline, representing inertial motion, and the non-geodesic
worldline of an accelerating particle. Since acceleration is absolute it
should be detectable (unlike the relative motion with constant velocity) and
it is the average anisotropic velocity of light in $N^{a}$ that constitutes
a way allowing an observer at rest in $N^{a}$ to detect the frame's
accelerated motion. That the propagation of light in non-inertial reference
frames (accelerating or at rest in a gravitational field) is anisotropic can
most clearly be demonstrated by revisiting the issue of propagation of light
in the Einstein elevator experiment \cite{infeld} and determining the
velocity of light rays parallel to the elevator's acceleration (in addition
to the horizontal ray originally considered by Einstein).

\begin{center}
\begin {picture}(0,110)(-40,130)
\setlength{\unitlength}{0.25mm}
\put(-100,100){\framebox(110,220){}} \put(10,212){\circle*{4}}
\put(-6,217){$B$} \put(29,212){\line(1,0){10}}
\put(34,201){\vector(0,1){11}} \put(10,190){\circle*{4}}
\put(-6,175){$B^{\prime}$} \put(29,190){\line(1,0){10}}
\put(34,201){\vector(0,-1){11}} \put(40,198){$\delta
=\frac{1}{2}at^{2}=\frac{ar^{2}}{2c^{2}}$} \put(-5,305){$A$}
\put(10,320){\circle*{4}} \put(10,100){\circle*{4}}
\put(-5,105){$C$} \put(-100.5,212){\circle*{4}} \put(-95,217){$D$}

\qbezier(-100.5,212)(-20,212)(2.5,193)
\put(2.5,193.3){\vector(3,-2){2}} \put(-121,96.6){---}
\put(-121,317.1){---} \put(-114,200){\vector(0,1){119.5}}
\put(-114,220.1){\vector(0,-1){119}} \put(-131,212){$2r$}
\put(-100,80){\line(0,1){10}} \put(10,80){\line(0,1){10}}
\put(-45,85){\vector(1,0){53.7}} \put(-45,85){\vector(-1,0){53.7}}
\put(-50,87.5){$r$} \put(17,320.5){\vector(0,-1){129}}
\put(17,99){\vector(0,1){89}} \thicklines
\put(-45,255){\vector(0,1){33}} \put(-39,267){{\bf a}}
\end {picture}
\end{center}

\vspace{2cm}

\begin{center}
\begin{list}{}{\leftmargin=1em \rightmargin=0em}\item[]
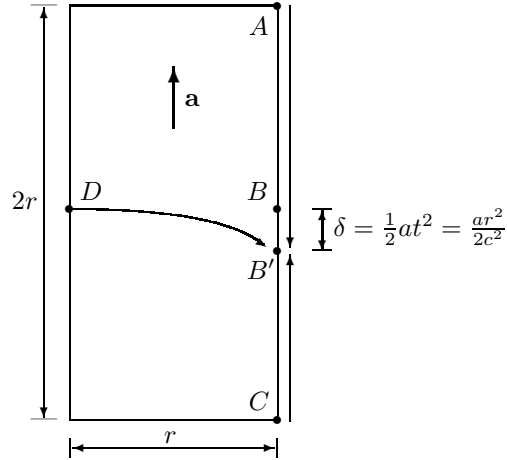
{\bf Figure 1}. Three light rays propagate in an accelerating
elevator. After having been emitted simultaneously from points
$A$, $C$, and $D$ the rays meet at $B^{\prime}$. The ray
propagating from $D$ towards $B$, but arriving at $B^{\prime}$,
represents the original thought experiment considered by Einstein.
The light rays emitted from $A$ and $C$ are introduced in order to
determine the expressions for the average velocity of light in an
accelerating frame of reference.
\end{list}
\end{center}

Consider an elevator accelerating with an acceleration $a=\left\vert \mathbf{%
a}\right\vert $ which represents a non-inertial (accelerating) reference
frame $N^{a}$ (Figure 1). Three light rays are emitted simultaneously in the
elevator (in $N^{a}$) from points $D$, $A$, and $C$ toward point $B$. Let $I$
be an inertial reference frame instantaneously at rest with respect to $%
N^{a} $ (i.e. the instantaneously comoving frame) at the moment the light
rays are emitted. The emission of the rays is therefore simultaneous in $%
N^{a}$ as well as in $I$ ($I$ and $N^{a}$ have a common instantaneous
three-dimensional space at this moment and therefore common simultaneity).
At the next moment an observer in $I$ sees that the three light rays arrive
simultaneously not at point $B$, but at $B^{\prime }$ since for the time $%
t=r/c$ the light rays travel toward $B$ the elevator moves at a distance $%
\delta =at^{2}/2=ar^{2}/2c^{2}$. As the simultaneous arrival of the three
rays at point $B^{\prime }$ as viewed in $I$ is an absolute
(observer-independent) fact due to its being a \emph{point} event, it
follows that the rays arrive simultaneously at $B^{\prime }$ as seen from $%
N^{a}$ as well. Since for the \emph{same} coordinate time $t=r/c$ in $N^{a}$
the three light rays travel \emph{different} distances $DB^{\prime }\approx
r $, $AB^{\prime }=r+\delta $, and $CB^{\prime }=r-\delta $, before arriving
simultaneously at point $B^{\prime }$, an observer in the elevator concludes
that the propagation of light is affected by the elevator's acceleration.
The \emph{average }velocity $c_{AB^{\prime }}^{a}$ of the light ray
propagating from $A$ to $B^{\prime }$ is slightly greater than $c$%
\[
c_{AB^{\prime }}^{a}=\frac{r+\delta }{t}\approx c\left( 1+\frac{ar}{2c^{2}}%
\right) .
\]%
The average velocity $c_{B^{\prime }C}^{a}$ of the light ray propagating
from $C$ to $B^{\prime }$ is slightly smaller than $c$%
\[
c_{CB^{\prime }}^{a}=\frac{r-\delta }{t}\approx c\left( 1-\frac{ar}{2c^{2}}%
\right) .
\]%
It is easily seen that to within terms $\sim c^{-2}$ the average light
velocity between $A$ and $B$ is equal to that between $A$ and $B^{\prime }$,
i.e. $c_{AB}^{a}=c_{AB^{\prime }}^{a}$ and also $c_{CB}^{a}=c_{CB^{\prime
}}^{a}$:
\begin{equation}
c_{AB}^{a}=\frac{r}{t-\delta /c}=\frac{r}{t-at^{2}/2c}=\frac{c}{1-ar/2c^{2}}%
\approx c\left( 1+\frac{ar}{2c^{2}}\right)  \label{c-ab}
\end{equation}%
and
\begin{equation}
c_{CB}^{a}=\frac{r}{t+\delta /c}\approx c\left( 1-\frac{ar}{2c^{2}}\right) .
\label{c-bc}
\end{equation}%
As the average velocities (\ref{c-ab}) and (\ref{c-bc}) are not determined
with respect to a specific point and since the \emph{coordinate} time $t$ is
involved in their calculation, it is clear that the expressions (\ref{c-ab})
and (\ref{c-bc}) represent the average \emph{coordinate} velocities between
the points $A$ and $B$ and $C$ and $B$, respectively.

The same expressions for the average coordinate velocities $c_{AB}^{a}$ and $%
c_{CB}^{a}$ can be also obtained from the expression for the coordinate
velocity of light in $N^{a}$. If the $z$-axis is parallel to the elevator's
acceleration $\mathbf{a}$ the spacetime metric in $N^{a}$ has the form \cite[%
p. 173]{misner}
\begin{equation}
ds^{2}=\left( 1+\frac{az}{c^{2}}\right) ^{2}c^{2}dt^{2}-dx^{2}-dy^{2}-dz^{2}.
\label{ds_a}
\end{equation}%
Note that due to the existence of a horizon at $z=-c^{2}/a$ \cite[pp. 169,
172-173]{misner} there are constrains on the size of non-inertial reference
frames (accelerated or at rest in a parallel gravitational field) which are
represented by the metric (\ref{ds_a}). If the origin of $N^{a}$ is changed,
say to $z_{B}=0$ (See Figure~1), the horizon moves to $z=-c^{2}/a-\left\vert
z_{B}\right\vert $.

As light propagates along null geodesics $ds^{2}=0$ the coordinate velocity
of light at a point $z$ in $N^{a}$ is
\begin{equation}
c^{a}(z)=\pm c\left( 1+\frac{az}{c^{2}}\right) .  \label{c_coord}
\end{equation}%
The $+$ and $-$ signs are for light propagating along or against $\ z$,
respectively. Therefore, the coordinate velocity of light at a point $z$ is
locally isotropic in the $z$ direction. It is clear that $c^{a}\left(
z\right) $ cannot become negative due to the constraints on the size of
non-inertial frames which ensure that $\left\vert z\right\vert <c^{2}/a$
\cite[pp. 169, 172]{misner}.

As the coordinate velocity $c^{a}\left( z\right) $ is continuous on the
interval $\left[ z_{A},\ z_{B}\right] $ one can calculate the average
coordinate velocity between $A$ and $B$ in Figure 1:

\begin{equation}
c_{AB}^{a}=\frac{1}{z_{B}-z_{A}}\int_{z_{A}}^{z_{B}}c^{a}\left( z\right)
dz=c\left( 1+\frac{az_{B}}{c^{2}}+\frac{ar}{2c^{2}}\right) ,  \label{c_AB}
\end{equation}%
where we took into account that $z_{A}=z_{B}+r$. When the coordinate origin
is at point $B$ ($z_{B}=0$) the expression (\ref{c_AB}) coincides with (\ref%
{c-ab}). In the same way:

\begin{equation}
c_{BC}^{a}=c\left( 1+\frac{az_{B}}{c^{2}}-\frac{ar}{2c^{2}}\right) ,
\label{c_BC}
\end{equation}%
where $z_{C}=z_{B}-r$. For $z_{B}=0$ (\ref{c_BC}) coincides with (\ref{c-bc}%
).

The average coordinate velocities (\ref{c_AB}) and (\ref{c_BC}) correctly
describe the propagation of light in $N^{a}$ yielding the right expression $%
\delta =ar^{2}/2c^{2}$ (See Figure~1). It should be stressed that without
these average coordinate velocities the fact that the light rays emitted
from $A$ and $C$ arrive not at $B,$ but at $B^{\prime }$ cannot be explained.

As a coordinate velocity, the average coordinate velocity of light is not
determined with respect to a specific point and depends on the choice of the
coordinate origin. Also, it is the same for light propagating from $A$ to $B$
and for light travelling in the opposite direction, i.e. $%
c_{AB}^{a}=c_{BA}^{a}$. Therefore, like the coordinate velocity (\ref%
{c_coord}) the average coordinate velocity is also isotropic but only in a
sense that the average light velocity between two points is the same in both
directions. As seen from (\ref{c_AB}) and (\ref{c_BC}) the average
coordinate velocity of light between different pairs of points, whose points
are the same distance apart, is different and this sense it is anisotropic.
As a result, as seen in Figure~1, the light ray emitted at $A$ arrives at $B$
before the light ray emitted at $C$.

In an elevator supported in a \emph{parallel} gravitational field
(representing a non-inertial reference frame $N^{g}$), where the metric is
\cite[p. 1056]{misner}

\begin{equation}
ds^{2}=\left( 1+\frac{2gz}{c^{2}}\right) c^{2}dt^{2}-dx^{2}-dy^{2}-dz^{2},
\label{ds_g}
\end{equation}%
the expressions for the average coordinate velocity of light between $A$ and
$B$ and $B$ and $C$, respectively, are

\[
c_{AB}^{g}=c\left( 1+\frac{gz_{B}}{c^{2}}+\frac{gr}{2c^{2}}\right)
\]
and

\[
c_{BC}^{g}=c\left( 1+\frac{gz_{B}}{c^{2}}-\frac{gr}{2c^{2}}\right) .
\]

The average coordinate velocity of light explains the propagation of light
in the Einstein elevator, but cannot be used in a situation where the
average light velocity between two points (say a source and an observation
point) is determined with respect to one of the points. Such situations
occur in the Shapiro time delay \cite{petkov2} and, as we shall see, when
one calculates the potential, the electric field, and the self-force of a
charge in non-inertial frames of reference. As the local velocity of light
is $c$ the average velocity of light between a source and an observation
point depends on which of the two points is regarded as a reference point
with respect to which the average velocity is determined (at the reference
point the local velocity of light is $c$). The dependence of the average
velocity on which point is chosen as a reference point demonstrates that
that velocity is anisotropic. This anisotropic velocity can be regarded as
an average\emph{\ proper} velocity of light since it is determined with
respect to a given point and its calculation involves the proper time at
that point.

The average proper velocity of light between $A$ and $B$ can be obtained by
using the average coordinate velocity of light (\ref{c_AB}) between the same
points:

\[
c_{AB}^{a}\equiv \frac{r}{\Delta t}=c\left( 1+\frac{az_{B}}{c^{2}}+\frac{ar}{%
2c^{2}}\right)
\]%
Let us calculated the average proper velocity of light propagating between $%
A $ and $B$ as determined from point $A.$ This means that we will use $A$'s
proper time $\Delta \tau _{A}=\left( 1+az_{A}/c^{2}\right) \Delta t$:

\[
\bar{c}_{AB}^{a}(as\ seen\ from\ A)\equiv \frac{r}{\Delta \tau _{A}}=\frac{r%
}{\Delta t}\frac{\Delta t}{\Delta \tau _{A}}.
\]%
Noting that $r/\Delta t$ is the average coordinate velocity (\ref{c_AB}) and
$z_{A}=z_{B}+r$ we have (to within terms $\sim c^{-2}$)

\begin{equation}
\bar{c}_{AB}^{a}(as\ seen\ from\ A)\approx c\left( 1+\frac{az_{B}}{c^{2}}+%
\frac{ar}{2c^{2}}\right) \left( 1-\frac{az_{A}}{c^{2}}\right) \approx
c\left( 1-\frac{ar}{2c^{2}}\right) .  \label{c-AB-A}
\end{equation}%
The calculation of the average proper velocity of light propagating between $%
A$ and $B$, but as seen from $B$ yields:

\begin{equation}
\bar{c}_{AB}^{a}(as\ seen\ from\ B)\equiv \frac{r}{\Delta \tau _{B}}=\frac{r%
}{\Delta t}\frac{\Delta t}{\Delta \tau _{B}}\approx c\left( 1+\frac{az_{B}}{%
c^{2}}+\frac{ar}{2c^{2}}\right) \left( 1-\frac{az_{B}}{c^{2}}\right) \approx
c\left( 1+\frac{ar}{2c^{2}}\right) .  \label{c-AB-B}
\end{equation}

The average proper velocities (\ref{c-AB-A}) and (\ref{c-AB-B}) can be
derived without using the average coordinate velocity of light. Consider a
light source at point $B$ (Figure 1). As seen from the metric (\ref{ds_a})
proper and coordinate times do not coincide whereas proper and coordinate
distances are the same \cite{rindler68}. To calculate the average proper
velocity of light originating from $B$ and observed at $A$ (that is, as seen
from $A$) we have to determine what are the initial velocity of a light
signal at $B$ and its final velocity at $A$ both with respect to $A$. As the
local velocity of light is $c$ the final velocity of the light signal
determined at $A$ is also $c$. Its initial velocity at $B$ as seen from $A$
is
\[
c_{B}^{a}=\frac{dz_{B}}{d\tau _{A}}=\frac{dz_{B}}{dt}\frac{dt}{d\tau _{A}}
\]%
where $dz_{B}/dt=c^{a}\left( z_{B}\right) $ is the coordinate velocity (\ref%
{c_coord}) at $B$%
\[
c^{a}\left( z_{B}\right) =c\left( 1+\frac{az_{B}}{c^{2}}\right)
\]%
and $d\tau _{A}$ is the proper time at $A$%
\[
d\tau _{A}=\left( 1+\frac{az_{A}}{c^{2}}\right) dt.
\]%
As $z_{A}=z_{B}+r$ and $az_{A}/c^{2}<1$ for the coordinate time $dt$ we have
(to within terms $\sim c^{-2}$)
\[
dt=\left( 1-\frac{az_{A}}{c^{2}}\right) d\tau _{A}=\left( 1-\frac{az_{B}}{%
c^{2}}-\frac{ar}{c^{2}}\right) d\tau _{A}.
\]%
Then for the initial velocity $c_{B}^{a}$ at $B$ as seen from $A$ we obtain
\[
c_{B}^{a}=c\left( 1+\frac{az_{B}}{c^{2}}\right) \left( 1-\frac{az_{B}}{c^{2}}%
-\frac{ar}{c^{2}}\right)
\]%
or keeping only the terms $\sim c^{-2}$
\begin{equation}
c_{B}^{a}=c\left( 1-\frac{ar}{c^{2}}\right) .  \label{c-proper}
\end{equation}%
The initial velocity $c_{B}^{a}$ at $B$ as seen from $A$ is the proper
velocity of light at $B$ determined by $A$. As seen from (\ref{c-proper})
this velocity is linear in both distance and time which allows us to
determine the average proper velocity of light between two points by using
only the proper velocities of light at those points. Therefore, for the
\emph{average proper} velocity $c_{BA}^{a}=(1/2)(c_{B}^{a}+c)$ of light
propagating from $B$ to $A$ \emph{as seen from} $A$ we have
\begin{equation}
c_{BA}^{a}\left( as\ seen\ from\ A\right) =c\left( 1-\frac{ar}{2c^{2}}%
\right) ,  \label{c_BA}
\end{equation}%
which coincides with (\ref{c-AB-A}).

As the local velocity of light at $A$ is $c$ it follows that if light
propagates from $A$ toward $B$ its average proper velocity $c_{AB}^{a}\left(
as\ seen\ from\ A\right) $ will be equal to the average proper velocity of
light propagating from $B$ toward $A$ $c_{BA}^{a}\left( as\ seen\ from\
A\right) $. Thus, as seen from $A$, the back and forth average proper
velocities of light travelling between $A$ and $B$ are the \emph{same}.

Now let us determine the average proper velocity of light between $B$ and $A$
with respect to the source point $B$. A light signal emitted at $B$ as seen
from $B$ will have an initial (local) velocity $c$ there. The final velocity
of the signal at $A$ as seen from $B$ will be
\[
c_{A}^{a}=\frac{dz_{A}}{d\tau _{B}}=\frac{dz_{A}}{dt}\frac{dt}{d\tau _{B}}
\]%
where $dz_{A}/dt=c^{a}\left( z_{A}\right) $ is the coordinate velocity at $A$%
\[
c^{a}\left( z_{A}\right) =c\left( 1+\frac{az_{A}}{c^{2}}\right)
\]%
and $d\tau _{B}$ is the proper time at $B$%
\[
d\tau _{B}=\left( 1+\frac{az_{B}}{c^{2}}\right) dt.
\]%
Then as $z_{A}=z_{B}+r$ we obtain for the velocity of light $c_{A}^{a}$ at $%
A $ as determined from\emph{\ }$B$
\[
c_{A}^{a}=c\left( 1+\frac{ar}{c^{2}}\right)
\]%
and the average proper velocity of light propagating from $B$ to $A$ \emph{%
as seen from }$B$ becomes
\begin{equation}
c_{BA}^{a}\left( as\ seen\ from\ B\right) =c\left( 1+\frac{ar}{2c^{2}}%
\right) .  \label{c-AB}
\end{equation}%
As expected (\ref{c-AB}) coincides with (\ref{c-AB-B}).

Comparing (\ref{c_BA}) and (\ref{c-AB}) demonstrates that the two average
proper velocities between the same points are not equal and depend on from
where they are seen. As we expected the fact that the local velocity of
light at the reference point is $c$ makes the average proper velocity
between two points dependant on where the reference point is.

In order to express the average proper velocity of light in a vector form
(which will turn out to be helpful for the calculation of the electric field
in the next section) let the light emitted from $B$ be observed at different
points. The average proper velocity of light emitted at $B$ and determined
at $A$ (Figure 1) is given by (\ref{c_BA}). As seen from point $C$ the
average proper velocity of light from $B$ to $C$ will be given by the same
expression as (\ref{c-AB})
\[
c_{BC}^{a}\left( as\ seen\ from\ C\right) =c\left( 1+\frac{ar}{2c^{2}}%
\right) .
\]%
As seen from a point $P$ at a distance $r$ from $B$ and lying on a line
forming an angle $\theta $ with the acceleration $\mathbf{a}$) the average
proper velocity of light from $B$ is
\[
c_{BP}^{a}\left( as\ seen\ from\ P\right) =c\left( 1-\frac{ar\cos \theta }{%
2c^{2}}\right) .
\]%
Then the average proper velocity of light coming from $B$ as seen from a
point defined by the position vector $\mathbf{r}$ originating from $B$ has
the form
\begin{equation}
\bar{c}^{a}=c\left( 1-\frac{\mathbf{a}\cdot \mathbf{r}}{2c^{2}}\right) .
\label{c_a}
\end{equation}

As evident from (\ref{c_a}) the average proper velocity of light emitted
from a common source and determined at different points around the source is
anisotropic in $N^{a}$ - if the observation point is above the light source
(with respect to the direction of $\mathbf{a}$) the average proper velocity
of light is slightly smaller than $c$\ and smaller than the average proper
velocity as determined from an observation point below the source. If an
observer at point $B$ (Figure 1) determines the average proper velocities of
light coming from $A$ and $C$ he will find that they are also anisotropic -
the average proper velocity of a light signal coming from $A$ is slightly
greater than that emitted at $C$ and therefore the $A$-signal will arrive
first. However, if the observer at $B$ (Figure 1) determines the average
proper velocities of light propagating between $A$ and $B$ and $B$ and $A$
he finds that the two average proper velocities are the same.

The calculation of the average proper velocity of light in the frame $N^{g}$
of an observer supported in a \emph{parallel} gravitational field can be
obtained in the same way and the resulting expression is:

\begin{equation}
\bar{c}^{g}=c\left( 1+\frac{\mathbf{g}\cdot \mathbf{r}}{2c^{2}}\right) ,
\label{c_g}
\end{equation}%
where $\mathbf{g}$ is the gravitational acceleration. It is interesting to
note that a substitution of $\mathbf{a=-g}$ in (\ref{c_a}) yields (\ref{c_g}%
) as required by the principle of equivalence. The next section will offer
us an additional reason to look into the question whether this result is an
indication that the identical anisotropic velocities of light in $N^{a}$ and
$N^{g}$ may provide us with an insight into what lies behind the equivalence
principle.

The velocities (\ref{c_a}) and (\ref{c_g}) demonstrate that there exists a
directional dependence in the propagation of light between two points in
non-inertial frames of reference (accelerating or at rest in a gravitational
field). This anisotropy in the propagation of light has been an overlooked
corollary of general relativity. In fact, up to now neither the average
coordinate velocity nor the average proper velocity of light have been
defined. However, we have seen that the average coordinate velocity is
needed to account for the propagation of light in non-inertial reference
frames (to explain the fact that two light signals emitted from points $A$,
and $C$ in Figure 1 meet at $B^{\prime }$, not at $B$). We will also see
below that the average proper velocity is necessary for the correct
description of electromagnetic phenomena in non-inertial frames in full
accordance with the equivalence principle.

\section{Electromagnetic mass theory and anisotropic velocity of light}

Here we shall not follow the standard approach to calculating the self-force
\cite{griffiths}, \cite{jackson}-\cite{podolsky} which describes an
electron's accelerated motion in an inertial frame $I$. Instead, all
calculations will be carried out in the non-inertial reference frame $N^{a}$
in which the accelerating electron is at rest. The reason for this is that
the calculation of the electric field and the self-force of an accelerating
electron in the accelerating frame $N^{a}$ (not in $I$) is crucial for the
correct application of the principle of equivalence since it relates those
quantities of an electron in a non-inertial (accelerating) frame $N^{a}$ and
in a non-inertial frame $N^{g}$ supported in a gravitational field. An
advantage of calculating the electron's electric field in the non-inertial
frame in which the electron is at rest is that it is obtained only from the
scalar potential and the calculations do not involve retarded times.

\subsection{An electron in a non-inertial (accelerating) reference frame $%
N^{a}$}

In the case of an accelerating reference frame the anisotropic velocity of
light (\ref{c_a}) leads to two changes in the potential
\begin{equation}
d\varphi ^{a}=-\frac{\rho dV^{a}}{4\pi \epsilon _{o}r^{a}}{,}
\label{pot_a_gen}
\end{equation}
of an electron at rest in $N^{a}$ and described there as compared to the
standard Coulomb potential of an inertial electron determined in its rest
frame, where $-\rho $ is the density of the electron charge, $r^{a}$ is the
distance between the charge element $-\rho dV^{a}$ and the observation
point, and $dV^{a}$ is an anisotropic volume element of the electron charge
(to be defined later) both determined in $N^{a}$. \noindent

First, analogously to determining the distance between a charge and an
observation point as $r=ct$ (where $t$ is the time it takes for an
electromagnetic signal to travel from the charge to the point at which the
potential is determined) in an inertial reference frame \cite[p. 416]%
{griffiths}, $r^{a}$ is expressed in $N^{a}$ as $r^{a}=\bar{c}^{a}t$. Since $%
\mathbf{a}\cdot \mathbf{r/}2c^{2}$ $<1$ we can write
\begin{equation}
\left( r^{a}\right) ^{-1}\approx r^{-1}\left( 1+\frac{\mathbf{a}\cdot
\mathbf{r}}{2c^{2}}\right) .  \label{r_a}
\end{equation}%
The second change in (\ref{pot_a_gen}) is a Li\'{e}nard-Wiechert-like (or
rather anisotropic) volume element $dV^{a}$ (not coinciding with the actual
volume element $dV$) which arises in $N^{a}$ on account of the anisotropic
velocity of light there. The origin of $dV^{a}$ is analogous to the origin
of the Li\'{e}nard-Wiechert volume element $dV^{LW}=dV/\left( 1-\mathbf{%
v\cdot n/}c\right) $ of a charge moving at velocity $\mathbf{v}$ with
respect to an inertial observer $I$, where $\mathbf{n=r/}r$ and $\mathbf{r}$
is the position vector at the retarded time \cite[p. 418]{griffiths}. The
origin of $dV^{LW}$ can be explained in terms of the \textquotedblright
information-collecting sphere\textquotedblright\ of Panofsky and Phillips
\cite[p. 342]{panofski} used in the derivation of the Li\'{e}nard-Wiechert
potentials; similar concepts are employed by Feynman \cite[p. 21-10]{feynman}%
, Griffiths \cite[p. 418]{griffiths}, and Schwartz \cite{schwartz}.

A charge approaching an observation point where the potential is determined
contributes more to the potential there ''than if the charge were
stationary'' \cite[p. 215]{schwartz} since it ''stays longer within the
information-collecting sphere'' \cite[p. 343]{panofski} which converges
toward the observation point and therefore chases the charge at the velocity
of light $c$ in $I$. The greater contribution to the potential may be viewed
as originating from a charge of a Li\'{e}nard-Wiechert volume $dV^{LW}$ that
appears greater than $dV$ \cite{charge}. If the charge is receding from the
observation point, the information-collecting sphere moves against the
charge, the charge stays less within it and the resulting smaller
contribution of the charge to the potential appears to come from a charge
whose Li\'{e}nard-Wiechert volume $dV^{LW}$ appears smaller than $dV$.

By the same argument the anisotropic volume $dV^{a}$ also appears different
from $dV$ in $N^{a}$. Consider a charge of length $l$ (placed in a direction
parallel to $\mathbf{a}$) at rest in $N^{a}$. The time for which the
information-collecting sphere traveling at the average velocity of light (%
\ref{c_a}) in $N^{a}$ sweeps over the charge is
\[
\Delta \tau ^{a}=\frac{l}{\bar{c}^{a}}=\frac{l}{c\left( 1-\mathbf{a}\cdot
\mathbf{r}/2c^{2}\right) }\approx \Delta t\left( 1+\frac{\mathbf{a}\cdot
\mathbf{r}}{2c^{2}}\right) ,
\]%
where $\Delta t=l/c$ is the time for which the information-collecting sphere
propagating at speed $c$ sweeps over an inertial charge of the same length $%
l $ in its rest frame. If the observation point where the potential is
determined is above the charge (in a direction parallel to $\mathbf{a}$),
the information-collecting sphere moves along $\mathbf{a}$ in $N^{a}$, its
average velocity is slightly smaller than $c$ (as seen from that point) and
therefore $\Delta \tau ^{a}>\Delta t$ (since $\mathbf{a}\cdot \mathbf{r}=ar$%
). As a result the charge stays longer within the sphere and its
contribution to the potential is greater. This is equivalent to saying that
the greater contribution comes from a charge of a greater length $l^{a}$
which for the same time $\Delta \tau ^{a}$ is swept over by an
information-collecting sphere propagating at velocity $c$:

\[
l^{a}=\Delta \tau ^{a}c=l\left( 1+\frac{\mathbf{a}\cdot \mathbf{r}}{2c^{2}}%
\right) .
\]%
If the point where the potential is determined is below the charge, the
information-collecting sphere moves against $\mathbf{a}$ in $N^{a}$ (meaning
that $\mathbf{a}\cdot \mathbf{r}=-ar$), its average velocity is slightly
greater than $c$ (as seen from the observation point) and the time for which
the sphere moves over the charge is smaller: $\Delta \tau ^{a}<\Delta t$.
Therefore the charge stays less time within the information-collecting
sphere and its contribution to the potential is smaller which can be
interpreted to mean that a charge of a shorter length is making a smaller
contribution to the potential.

The anisotropic volume element which corresponds to such an apparent length $%
l^{a}$ is obviously
\begin{equation}
dV^{a}=dV\left( 1+\frac{\mathbf{a}\cdot \mathbf{r}}{2c^{2}}\right) .
\label{dV_a}
\end{equation}

The scalar potential of a charged volume element $-\rho dV^{a}$ of the
electron at rest in $N^{a}$ can now be calculated by substituting (\ref{r_a}%
) and (\ref{dV_a}) in (\ref{pot_a_gen})
\[
d\varphi ^{a}=-\frac{1}{4\pi \epsilon _{0}}\frac{\rho dV^{a}}{r^{a}}=-\frac{1%
}{4\pi \epsilon _{0}}\frac{\rho dV}{r}\left( 1+\frac{\mathbf{a}\cdot \mathbf{%
r}}{2c^{2}}\right) ^{2}
\]%
or keeping only the terms proportional to $c^{-2}$ we obtain
\begin{equation}
d\varphi ^{a}=-\frac{\rho }{4\pi \epsilon _{0}r}\left( 1+\frac{\mathbf{a}%
\cdot \mathbf{r}}{c^{2}}\right) dV.  \label{pot_a}
\end{equation}%
The electric field of the charged volume element $-\rho dV^{a}$ can be
calculated only from the scalar potential (\ref{pot_a}) without the
involvement of retarded times since the charged element is at rest in $N^{a}$
\[
d\mathbf{E}^{a}=-\nabla d\varphi ^{a}=-\frac{1}{4\pi \epsilon _{o}}\left(
\frac{\mathbf{n}}{r^{2}}+\frac{\mathbf{a\cdot n}}{c^{2}r}\mathbf{n}-\frac{%
\mathbf{a}}{c^{2}r}\right) \rho dV.
\]%
The electric field of the electron then is
\begin{equation}
\mathbf{E}^{a}=-\frac{1}{4\pi \epsilon _{o}}\int \left( \frac{\mathbf{n}}{%
r^{2}}+\frac{\mathbf{a}\cdot \mathbf{n}}{c^{2}r}\mathbf{n}-\frac{\mathbf{a}}{%
c^{2}r}\right) \rho dV.  \label{E_a}
\end{equation}

If we compare the electric field (\ref{E_a}) of an electron at rest in $%
N^{a} $(determined in $N^{a}$) and its field \cite[p. 664]{jackson}
determined in an inertial reference frame $I$ in which the electron is
instantaneously at rest we see that for both an observer in $N^{a}$ and an
observer in $I$ the electron's field is equally distorted. Therefore, as we
expected it does turn out that the shape of the electric field of an
accelerating charge is absolute like acceleration itself. This implies that
there exists a correspondence between the shape of a charge and its state of
motion which is observer-independent. As for all observers (both inertial
and non-inertial) a worldline is either geodesic or not, the field of an
inertial charge represented by a geodesic worldline is the Coulomb field,
whereas the field of a non-inertial charge (accelerating or, as we shall see
below, supported in a gravitational field) whose worldline is not geodesic
is distorted.

The self-force which the field of the electron exerts upon an element $-\rho
dV_{1}^{a}$ of its own charge is
\begin{equation}
d\mathbf{F}_{self}^{a}=-\rho dV_{1}^{a}\mathbf{E}^{a}=\frac{1}{4\pi \epsilon
_{o}}\int \left( \frac{\mathbf{n}}{r^{2}}+\frac{\mathbf{a}\cdot \mathbf{n}}{%
c^{2}r}\mathbf{n}-\frac{\mathbf{a}}{c^{2}r}\right) \rho ^{2}dVdV_{1}^{a}{.}
\label{dF_a}
\end{equation}
The resultant self-force acting on the electron as a whole is:
\begin{equation}
\mathbf{F}_{self}^{a}=\frac{1}{4\pi \epsilon _{o}}\int \int \left( \frac{%
\mathbf{n}}{r^{2}}+\frac{\mathbf{a}\cdot \mathbf{n}}{c^{2}r}\mathbf{n}-\frac{%
\mathbf{a}}{c^{2}r}\right) \rho ^{2}dVdV_{1}^{a}{,}  \label{F_a1}
\end{equation}

\noindent which after taking into account the anisotropic volume element (%
\ref{dV_a}) becomes
\[
\mathbf{F}_{self}^{a}=\frac{1}{4\pi \epsilon _{o}}\int \int \left( \frac{%
\mathbf{n}}{r^{2}}+\frac{\mathbf{a}\cdot \mathbf{n}}{c^{2}r}\mathbf{n}-\frac{%
\mathbf{a}}{c^{2}r}\right) \left( 1+\frac{\mathbf{a}\cdot \mathbf{r}}{2c^{2}}%
\right) \rho ^{2}dVdV_{1}{.}
\]

Assuming a spherically symmetric distribution of the electron charge \cite%
{lorentz} and following the standard procedure of calculating the self-force
\cite{podolsky} we get (see Appendix):
\begin{equation}
\mathbf{F}_{self}^{a}=-\frac{U}{c^{2}}\mathbf{a}{,}  \label{F_a}
\end{equation}
where
\[
U=\frac{1}{8\pi \epsilon _{o}}\int \int \frac{\rho ^{2}}{r}dVdV_{1}
\]

\noindent is the energy of the electron's electric field. As $U/c^{2}$ is
the mass that corresponds to that energy we can write (\ref{F_a}) in the
form:
\begin{equation}
\mathbf{F}_{self}^{a}=-m^{a}\mathbf{a,}  \label{F=ma}
\end{equation}

\noindent where $m^{a}=U/c^{2}$ is identified as the electron inertial
electromagnetic mass. The famous factor of $4/3$ in the electromagnetic mass
of the electron does not appear in (\ref{F=ma}). The reason is that in (\ref%
{dF_a}) and (\ref{F_a1}) we have identified and used the correct volume
element $dV_{1}^{a}=\left( 1+\mathbf{a}\cdot \mathbf{r/}2c^{2}\right) dV_{1}$
originating from the anisotropic velocity of light in $N^{a}$; not taking it
into account results in the appearance of the $4/3$ factor.

The self-force $\mathbf{F}_{self}^{a}$ to which an electron is subjected due
to its own distorted field is directed opposite to $\mathbf{a}$ and
therefore resists the acceleration of the electron. As seen from (\ref{F_a})
this force is purely electromagnetic in origin and therefore both the
resistance the classical electron offers to being accelerated (i.e. its
inertia) and its inertial mass (which is a measure of that resistance) are
purely electromagnetic in origin as well.

The self-force (\ref{F=ma}) is traditionally called the inertial force.
According to Newton's third law the external force $\mathbf{F}$ that
accelerates the electron and the self-force $\mathbf{F}_{self}^{a}$ which
resists $\mathbf{F}$ have equal magnitudes and opposite directions: $\mathbf{%
F=-F}_{self}^{a}$. Therefore we can write $\mathbf{F}=m^{a}\mathbf{a}$ which
means that Newton's second law can be \emph{derived} on the basis of
Maxwell's electrodynamics applied to the classical model of the electron and
Newton's third law.

We have seen that the average anisotropic velocity of light in $N^{a}$
causes the imbalance in the \emph{repulsion} of all volume elements $-\rho
dV^{a}$ and $-\rho dV_{1}^{a}$ that is responsible for the self-force $%
\mathbf{F}_{self}^{a}$ to which an electron as a whole is subjected and
which resists its accelerated motion. While it is the net effect of the
unbalanced repulsion of the electron volume elements that gives rise to its
electromagnetic mass, the contribution to the mass of the unbalanced
repulsion of individual charges is more complex. The unbalanced repulsion of
two like charges whose line of interaction is perpendicular to the
acceleration $\mathbf{a}$ increases their mass. The unbalanced repulsion of
two charges whose line of interaction is parallel to $\mathbf{a}$, however,
decreases their mass as can be verified by direct calculation. By taking
into account the anisotropic volume element $dV_{1}^{a}$ in (\ref{dF_a}) the
latter effect has resulted in reducing the electron mass from $4/3\ m$ to $m$%
.

If the unbalanced repulsion of like charges causes the classical electron's
inertia, mass, and the self-force $\mathbf{F}_{self}^{a}$ that acts in a
direction opposite to $\mathbf{a}$, it is quite natural to ask what the
effect of the unbalanced \emph{attraction} of two unlike charges $-\rho
dV^{a}$ and $+\rho dV_{1}^{a}$ will be. As in the case of unbalanced
repulsion it depends on the angle between $\mathbf{a}$ and the line of
interaction of the two opposite charges. The unbalanced attraction of two
charges $-\rho dV^{a}$ and $+\rho dV_{1}^{a}$ whose line of interaction is
perpendicular to the acceleration $\mathbf{a}$ \emph{decreases} their mass
by $dm$ since the self-force acting on them is in the direction of $\mathbf{a%
}$: $d\mathbf{F}_{self}^{a}=dm\mathbf{a}$. This is an inevitable but
nevertheless a surprising result since it demonstrates that not only does
the self-force acting on accelerating opposite charges not resist their
acceleration but further increases it \cite{cornish}, \cite{petkov3} (note
that this effect is different from the runaway problem \cite{coleman} in the
classical electron theory). On the other hand, the negative contribution to
the mass resulting from unbalanced attraction is not so surprising - in
Section 3 we have seen that non-electric \emph{attraction} forces also make
a negative contribution to the mass. When the line of interaction of two
opposite charges is parallel to $\mathbf{a}$ their unbalanced attraction
increases their mass since the self-force in this case is $d\mathbf{F}%
_{self}^{a}=-\ dm\mathbf{a}$.

Let us now calculate the electric field of an inertial electron whose charge
is $e$ and which appears falling in $N^{a}$ with an apparent acceleration $%
\mathbf{a}^{\ast }=-\mathbf{a}$ (where $\mathbf{a}$ is the acceleration of $%
N^{a}$). It is obvious that for an inertial observer $I$ falling with the
electron its electric field is the Coulomb field. In order to obtain the
electric field of the falling electron in $N^{a}$ we have to use generalized
Li\'{e}nard-Wiechert potentials which take into account the anisotropic
velocity of light in $N^{a}$. These can be directly obtained by replacing $r$
with $r^{a}$ and the actual volume element $dV$ with the anisotropic volume
element $dV^{a}$ in the expressions for the Li\'{e}nard-Wiechert potentials
in an inertial reference frame:
\begin{equation}
\varphi ^{a}\left( r,t\right) =\left\vert -\frac{e}{4\pi \epsilon _{o}r}%
\frac{1}{1-\mathbf{v\cdot n/}c}\left( 1+\frac{\mathbf{a}\cdot \mathbf{r}}{%
c^{2}}\right) \right\vert _{ret}  \label{LWa_s}
\end{equation}%
\begin{equation}
\mathbf{A}^{a}\left( r,t\right) =\left\vert -\frac{e}{4\pi \epsilon
_{o}c^{2}r}\frac{\mathbf{v}}{1-\mathbf{v\cdot n/}c}\left( 1+\frac{\mathbf{a}%
\cdot \mathbf{r}}{c^{2}}\right) \right\vert _{ret},  \label{LWa_v}
\end{equation}%
where, as usual, the subscript \textquotedblright ret\textquotedblright\
indicates that the quantity in the square brackets is evaluated at the
retarded time. The electric field of an electron falling in $N^{a}$ (and
considered instantaneously at rest in $N^{a}$) obtained from (\ref{LWa_s})
and (\ref{LWa_v}) is:
\[
\mathbf{E}=-\nabla \varphi ^{a}-\frac{\partial \mathbf{A}^{a}}{\partial t}=-%
\frac{e}{4\pi \epsilon _{o}}\left[ \left( \frac{\mathbf{n}}{r^{2}}+\frac{%
\mathbf{a}^{\ast }\mathbf{\cdot n}}{c^{2}r}\mathbf{n}-\frac{\mathbf{a}^{\ast
}}{c^{2}r}\right) +\left( \frac{\mathbf{a}\cdot \mathbf{n}}{c^{2}r}\mathbf{n}%
-\frac{\mathbf{a}}{c^{2}r}\right) \right] .
\]

\noindent Noting that $\mathbf{a}^{\ast }=-\mathbf{a}$ it proves that the
electric field of the falling electron, as described in $N^{a}$, is
identical with the field of an inertial electron determined in its rest
frame:
\[
\mathbf{E}=-\frac{e}{4\pi \epsilon _{o}}\frac{\mathbf{n}}{r^{2}}.
\]%
This result shows that for a non-inertial observer at rest in $N^{a}$ the
instantaneous electric field of the falling electron is the Coulomb field.
Therefore, the assumption that the shape of the electric field of an
inertial electron is absolute is also confirmed since both $I$ and $N^{a}$
detect a Coulomb field of the falling electron. In general: (i) a Coulomb
field is associated with an inertial electron (represented by a geodesic
worldline) by both an inertial observer $I$ (moving with the electron) and a
non-inertial observer $N^{a}$, and (ii) for both $I$ and $N^{a}$ the
electric field of a non-inertial electron (whose worldline is not geodesic)
is equally distorted. As we expected the fact that the state of (inertial or
accelerated) motion of a charge is absolute implies that the shape of the
electric field of an (inertial or accelerated) charge is also absolute (the
same for an inertial and a non-inertial observer).

\subsection{An electron in a non-inertial reference frame $N^{g}$ at rest in
the Earth's gravitational field}

Similarly to the case of calculating the electric potential in $N^{a}$ the
average anisotropic velocity of light (\ref{c_g}) in $N^{g}$ also leads to
anisotropic $r^{g}$ and $dV^{g}$ in $N^{g}$:
\[
\left( r^{g}\right) ^{-1}\approx r^{-1}\left( 1-\frac{\mathbf{g}\cdot
\mathbf{r}}{2c^{2}}\right)
\]%
and
\begin{equation}
dV^{g}=dV\left( 1-\frac{\mathbf{g}\cdot \mathbf{r}}{2c^{2}}\right) .
\label{dV_g}
\end{equation}%
As a result the scalar potential of a charged volume element $-\rho dV^{g}$
of the electron in $N^{g}$ is:
\begin{equation}
d\varphi ^{g}=-\frac{\rho }{4\pi \epsilon _{0}r}\left( 1-\frac{\mathbf{g}%
\cdot \mathbf{r}}{c^{2}}\right) dV.  \label{pot_g}
\end{equation}%
It should be emphasized that only by taking into account the anisotropic
volume element $dV^{g}$ we can obtain the correct potential (\ref{pot_g}) of
a charge supported in a gravitational field. By using the actual volume
element $dV$ in 1921 Fermi \cite{fermi21} derived the expression
\begin{equation}
\varphi =\frac{e}{4\pi \epsilon _{0}r}\left( 1-\frac{1}{2}\frac{gz}{c^{2}}%
\right)  \label{1fermi}
\end{equation}%
for the potential of a charge at rest in a gravitational field. The use of $%
dV$ resulted in the factor $1/2$ in the parenthesis which leads to a
contradiction with the principle of equivalence when the electric field is
calculated from this potential: it follows from (\ref{1fermi}) that the
electric field of a charge supported in the Earth's gravitational field
coincides with the instantaneous electric field of a charge moving with an
acceleration $\mathbf{a}=-\mathbf{g}/2$ (obviously the principle of
equivalence requires that $\mathbf{a}=-\mathbf{g}$).

The calculation of the electric field of a charged volume element $-\rho
dV^{g}$ in $N^{g}$ is again carried out by using only the scalar potential (%
\ref{pot_g}):
\[
d\mathbf{E}^{g}=-\nabla d\varphi ^{g}=-\frac{1}{4\pi \epsilon _{o}}\left(
\frac{\mathbf{n}}{r^{2}}-\frac{\mathbf{g\cdot n}}{c^{2}r}\mathbf{n}+\frac{%
\mathbf{g}}{c^{2}r}\right) \rho dV
\]

\noindent and the field of the electron is then
\begin{equation}
\mathbf{E}^{g}=-\frac{1}{4\pi \epsilon _{o}}\int \left( \frac{\mathbf{n}}{%
r^{2}}-\frac{\mathbf{g\cdot n}}{c^{2}r}\mathbf{n}+\frac{\mathbf{g}}{c^{2}r}%
\right) \rho dV{.}  \label{E_g}
\end{equation}%
A comparison of the electric field of an electron supported in the Earth's
gravitational field (\ref{E_g}), determined in $N^{g}$, with the electric
field of an accelerated electron (\ref{E_a}), determined in the frame $N^{a}$%
, indicates that the electric fields of an electron at rest on the Earth's
surface and an electron at rest in the frame $N^{a}$ which moves with an
acceleration $\mathbf{a}=-\mathbf{g}$ are equally distorted in accordance
with the principle of equivalence. A substitution $\mathbf{a}=-\mathbf{g}$
in (\ref{pot_a}) also transforms it into (\ref{pot_g}) as required by the
equivalence principle.

The self-force with which the electron field interacts with an element $%
-\rho dV_{1}^{g}$ of the electron charge is therefore
\begin{equation}
d\mathbf{F}_{self}^{g}=-\rho dV_{1}^{g}\mathbf{E}^{g}=\frac{1}{4\pi \epsilon
_{o}}\int \left( \frac{\mathbf{n}}{r^{2}}-\frac{\mathbf{g}\cdot \mathbf{n}}{%
c^{2}r}\mathbf{n}+\frac{\mathbf{g}}{c^{2}r}\right) \rho ^{2}dVdV_{1}^{g}
\label{dF_g}
\end{equation}
and the resultant self-force with which the electron acts upon itself is:
\begin{equation}
\mathbf{F}_{self}^{g}=\frac{1}{4\pi \epsilon _{o}}\int \int \left( \frac{%
\mathbf{n}}{r^{2}}-\frac{\mathbf{g}\cdot \mathbf{n}}{c^{2}r}\mathbf{n}+\frac{%
\mathbf{g}}{c^{2}r}\right) \rho ^{2}dVdV_{1}^{g}{,}  \label{F_g}
\end{equation}

\noindent After taking into account the explicit form (\ref{dV_g}) of $%
dV_{1}^{g}$, assuming a spherically symmetric distribution of the electron
charge, and calculating the self-force as we have done in the case of an
electron at rest in $N^{a}$ we get:
\begin{equation}
\mathbf{F}_{self}^{g}=\frac{U}{c^{2}}\mathbf{g},  \label{F_g1}
\end{equation}
where

\[
U=\frac{1}{8\pi \epsilon _{o}}\int \int \frac{\rho ^{2}}{r}dVdV_{1}
\]

\noindent is the energy of the electron's field. As $U/c^{2}$ is the mass
associated with the field energy of the electron, i.e. its electromagnetic
mass, (\ref{F_g1}) obtains the form:
\begin{equation}
\mathbf{F}_{self}^{g}=m^{g}\mathbf{g,}  \label{F=mg}
\end{equation}

\noindent where $m^{g}=U/c^{2}$ is interpreted here as the electron passive
gravitational mass. As in the case of the self-force acting on an
accelerating electron described in $N^{a}$ the $4/3$ factor in the
electromagnetic mass does not appear in (\ref{F=mg}) for the same reason:
the correct volume element (\ref{dV_g}) was used in (\ref{F_g}). Therefore
the anisotropic volume element $dV^{g}$ \emph{simultaneously} resolves two
different problems - removes both the $1/2$ factor in the potential (\ref%
{1fermi}) derived by Fermi and the $4/3$ factor in the self-force.

The self-force $\mathbf{F}_{self}^{g}$ which acts upon an electron on
account of its own distorted field is directed parallel to $\mathbf{g}$ and
resists the deformation of its electric field caused by the fact that the
electron at rest on the Earth's surface is prevented from falling. This
force is traditionally called the gravitational force. As we have seen $%
\mathbf{F}_{self}^{g}$ arises only when an electron is prevented from
falling, i.e. only when it is deviated from its geodesic path. Only in this
case the electron field deforms which gives rise to the self-force $\mathbf{F%
}_{self}^{g}$. Thus $\mathbf{F}_{self}^{g}$ resists the deformation of the
field of the electron which means that it resists its being prevented from
following a geodesic path. As a Coulomb field is associated with a
non-resistantly moving electron (represented by a geodesic worldline) it
follows that $\mathbf{F}_{self}^{g}$ is, in fact, an \emph{inertial} force
since it resists the deviation of an electron from its geodesic path, that
is, $\mathbf{F}_{self}^{g}$ resists the deviation of the electron from its
motion by inertia. Therefore, the nature of the force acting upon an
electron at rest in a gravitational field is inertial and is purely
electromagnetic in origin as seen from (\ref{F=mg}), which means that the
electron passive gravitational mass $m^{g}$ in $\mathbf{F}_{self}^{g}=m^{g}%
\mathbf{g}$ is also purely electromagnetic in origin. It is immediately
clear from here why the inertial and the passive gravitational masses of the
classical electron are equal. As the nature of the self-force $\mathbf{F}%
_{self}^{g}$ is inertial it follows that what is traditionally called
passive gravitational mass is, in fact, also inertial mass. This becomes
evident from the fact that the two masses are the measure of resistance an
electron offers when deviated from its geodesic path. In flat spacetime the
force of resistance is $\mathbf{F}_{self}^{a}=-m^{a}\mathbf{a}$ whereas in
curved spacetime this same force that resists the deviation of the electron
from its geodesic path is $\mathbf{F}_{self}^{g}=m^{g}\mathbf{g}$ where $%
m^{a}$ and $m^{g}$ are the measures of resistance (inertia) in both cases.
The two resistance force are equal $F_{self}^{a}=F_{self}^{g}$ for $a=g$ as
seen from (\ref{F_a}) and (\ref{F_g1}) and therefore $m^{a}=m^{g}$. This
equivalence also follows from the fact that $m^{a}$ and $m^{g}$ are the
\emph{same} thing - the mass associated with the energy of the electron
field.

The result that the force $\mathbf{F}_{self}^{g}$, acting on an electron
when it is deviated from its geodesic path due to its being at rest in a
gravitational field, is inertial is valid not only for the classical
electron. A non-resistant motion (i.e. motion by inertia) of a body in both
special relativity (flat spacetime) and general relativity (curved
spacetime) is represented by a geodesic worldline whereas a body represented
by a non-geodesic worldline is subjected to a resistance force which opposes
the external force preventing the body from following its geodesic path in
spacetime. That is why the nature of the resistance force is inertial in
both special and general relativity. It should be specifically stressed here
that the conclusion of the non-gravitational nature of the force acting on a
body at rest in a gravitational field follows from general relativity itself
\cite[p. 244]{rindler}: as a body supported in a gravitational field is
deviated from its geodesic path, which means that it is prevented from
moving non-resistantly (by inertia), it is subjected to an inertial force
since it arises only when the body is prevented from moving in a
non-resistant (inertial) manner. This explains why in general relativity
\textquotedblright there is no such thing as the force of
gravity\textquotedblright\ \cite{synge}.

As seen from (\ref{dF_g}) it is the unbalanced \emph{repulsion} of all
volume elements $-\rho dV^{g}$ and $-\rho dV_{1}^{g}$ (caused by the average
anisotropic velocity of light in $N^{g}$) that is responsible for the
self-force $\mathbf{F}_{self}^{g}$ to which the electron as a whole is
subjected (if the velocity of light were isotropic, all forces of repulsion
of the electron volume elements would cancel out and there would be no
self-force acting upon the electron as a whole). As in the acceleration case
here too the unbalanced \emph{attraction} of two unlike charges $-\rho dV^{g}
$ and $+\rho dV_{1}^{g}$ changes the sign in the self-force in (\ref{dF_g}),
which means that for the case when the line of interaction of the charges is
perpendicular to $\mathbf{g}$ it becomes a levitation force \cite%
{griffiths86}, \cite{petkov3} whose effect is to reduce the charges' passive
gravitational electromagnetic mass. The unbalanced attraction of unlike
charges whose line of interaction is parallel to $\mathbf{g}$ results in a
self-force in the direction of $\mathbf{g}$ which increases their mass. In a
gravitational field, the unbalanced repulsion of two like charges
interacting along a line perpendicular to $\mathbf{g}$ produces a self-force
parallel to $\mathbf{g}$. The self-force produced by the unbalanced
repulsion of the like charges interacting along a line parallel to $\mathbf{g%
}$ is opposite to $\mathbf{g}$.

Let us now calculate the electric field of an electron falling in the
Earth's gravitational field. General relativity describes an electron
falling in a gravitational field by a geodesic worldline. It implies that it
moves by inertia (without resistance) and its Coulomb field is not distorted
for an inertial observer $I$ falling with the electron. In order to obtain
the electric field of an electron falling with acceleration $\mathbf{a}=%
\mathbf{g}$ in the Earth's gravitational field (in $N^{g}$) we have to use
the generalized Li\'{e}nard-Wiechert potentials which include the
corrections due to the average anisotropic velocity of light in $N^{g}$ as
we have done in the case of an electron falling in $N^{a}$. Their explicit
expressions in $N^{g}$ are:
\begin{equation}
\varphi ^{g}\left( r,t\right) =\left\vert -\frac{e}{4\pi \epsilon _{o}r}%
\frac{1}{1-\mathbf{v\cdot n}/c}\left( 1-\frac{\mathbf{g}\cdot \mathbf{r}}{%
c^{2}}\right) \right\vert _{ret}  \label{LWs_g}
\end{equation}%
\begin{equation}
\mathbf{A}^{g}\left( r,t\right) =\left\vert -\frac{e}{4\pi \epsilon
_{o}c^{2}r}\frac{\mathbf{v}}{1-\mathbf{v\cdot n}/c}\left( 1-\frac{\mathbf{g}%
\cdot \mathbf{r}}{c^{2}}\right) \right\vert _{ret}.  \label{LWv_g}
\end{equation}%
The electric field of the electron falling in $N^{g}$ (and considered
instantaneously at rest in $N^{g}$) obtained from (\ref{LWs_g}) and (\ref%
{LWv_g}) is:
\[
\mathbf{E}=-\nabla \varphi ^{g}-\frac{\partial \mathbf{A}^{g}}{\partial t}=-%
\frac{e}{4\pi \epsilon _{o}}\left[ \left( \frac{\mathbf{n}}{r^{2}}+\frac{%
\mathbf{g\cdot n}}{c^{2}r}\mathbf{n}-\frac{\mathbf{g}}{c^{2}r}\right)
+\left( -\frac{\mathbf{g}\cdot \mathbf{n}}{c^{2}r}\mathbf{n}+\frac{\mathbf{g}%
}{c^{2}r}\right) \right] .
\]

\noindent Therefore the electric field of the falling electron in the
reference frame $N^{g}$ proves to be identical with the field of an inertial
electron determined in its rest frame:
\begin{equation}
\mathbf{E}=-\frac{e}{4\pi \epsilon _{o}}\frac{\mathbf{n}}{r^{2}}.
\label{E_in}
\end{equation}

As seen from (\ref{E_in}) while the electron is falling in the Earth's
gravitational field its electric field at any instant is the Coulomb field
which means that no self-force is acting on the electron, i.e. there is no
resistance to its accelerated motion. This explains why the electron is
falling in a gravitational field \textit{by itself} and no external force is
causing its acceleration. As (\ref{E_in}) shows, the only way for the
electron to compensate the anisotropy in the propagation of light and to
preserve the Coulomb shape of its electric field is to fall with an
acceleration $\mathbf{g}$. If the electron is prevented from falling its
electric field distorts, the self-force (\ref{F=mg}) appears and tries to
force the electron to move (fall) in such a way that its field becomes the
Coulomb field; as a result the self-force disappears.

The result that the electron (and any charge) falls in a gravitational field
with an acceleration $\mathbf{g}$ by itself in order to prevent its field
from getting distorted may shed light on two facts: (i) that in general
relativity the motion of a body falling toward a gravitating center is
regarded as inertial (non-resistant) and is represented by a geodesic
worldline, and (ii) the experimental fact that all objects fall in a
gravitational field with the \emph{same} acceleration.

The conclusion that the electron preserves the Coulomb shape of its field
while falling in a gravitational field is also quite relevant to the debate
\cite{rohrlich90}, \cite{dewitt}-\cite{harpaz} on whether or not a falling
charge radiates. As (\ref{E_in}) shows the electric field of a falling
charge is the Coulomb field and therefore does not contain the radiation $%
r^{-1}$ terms - this is a straightforward demonstration that a charge
falling in a gravitational field does not radiate \cite{petkov}.

The result (\ref{E_in}) demonstrates the important fact that a Coulomb field
is associated with the falling electron by \emph{both} an inertial observer $%
I$ (falling with the electron) and a non-inertial observer at rest in $N^{g}$%
. This again confirms the assumption that the shape of the electric field of
an inertial electron (represented by a geodesic worldline) is absolute
(observer-independent) due to the fact that the inertial motion is itself
absolute.

Now we are in a position to answer all four questions formulated in Section
3. (i) An electron moving with constant velocity in a flat spacetime region
where the average velocity of light is isotropic does not resist its uniform
motion since uniform motion (represented by a straight geodesic worldline)
in flat spacetime ensures that the electron's electric field is the Coulomb
field. Stated another way, the only way for an electron to prevent its
electric field from distorting in flat spacetime is to move with constant
velocity. (ii) An accelerating electron resists its acceleration in flat
spacetime because the accelerated motion distorts the electron's electric
field which results in an electric self-force that opposes the deformation
of the electron's field. (iii) An electron falling toward the Earth's
surface does not resists its (flat-spacetime) acceleration since, as we have
seen, falling with an acceleration $\mathbf{g}$ is the only way for the
electron to compensate the anisotropy in the propagation of light in the
Earth's vicinity and to prevent its electric field from getting distorted
(the curved-spacetime acceleration of the falling electron is zero). (iv) An
electron at rest on the Earth's surface is subjected to an electric
self-force trying to make the electron fall since the average anisotropic
velocity of light in the Earth's gravitational field distorts the electron
field which in turn gives rise to the self-force. The nature of that force
is inertial (not gravitational) and is electromagnetic in origin.

The mechanism giving rise to the free (non-resistant) fall of the electron
in a gravitational field and to the self-force (\ref{F=mg}) that resists its
prevention from falling is identical to the mechanism responsible for the
self-force (\ref{F_a}) an accelerated electron in flat spacetime is
subjected to and for its free fall as described in the accelerated frame $%
N^{a}$. This mechanism implies that the common anisotropy in the propagation
of light in $N^{a}$ and $N^{g}$ gives rise to similar phenomena whose
mathematical expressions transform into one another when the substitution $%
\mathbf{a}=-\mathbf{g}$ is used. Therefore, the equivalence principle seems
to originate from the same metrics (\ref{ds_a}) and (\ref{ds_g}) which give
rise to the anisotropic propagation of light in the non-inertial frames of
reference $N^{a}$ and $N^{g}$ and to \emph{identical} electromagnetic
phenomena in $N^{a}$ and $N^{g}$.

\subsection{Gravitational interaction of the classical electron}

We have shown that the inertial and passive gravitational masses of the
classical electron are entirely electromagnetic in origin. As all three
masses - inertial, passive gravitational, and active gravitational - are
considered equal \cite{agravmass2}, it follows that the electron active
gravitational mass is fully electromagnetic in origin as well. And since it
is only the charge of the classical electron that represents it (there is no
mechanical mass), it follows that the active gravitational mass of the
electron is represented by its charge. Therefore it is the electron charge
that distorts spacetime and causes the average anisotropic velocity of light
in the electron's neighborhood.

We have seen that it is the anisotropy in the average proper velocity of
light and the electromagnetic mass theory that fully and consistently
explain the fall of an electron toward the Earth and the self-force acting
on an electron at rest on the Earth's surface. Let us now see whether the
gravitational attraction between two electrons can be explained in the same
way.

In addition to the electric repulsion of two electrons ($e_{1}$ and $e_{2}$)
in space, they also attract each other through the anisotropy in the average
proper velocity of light caused by each of them: $e_{1}$ falls toward $e_{2}$
in order to compensate the anisotropy caused by $e_{2}$ and to prevent its
electric field from getting distorted and vice versa. In other words, the
charge of the electrons affects the propagation of light around them which
in turn changes the shape of the electrons' worldlines resulting in their
convergence toward each other. Therefore, in the framework of the
electromagnetic mass theory the anisotropy in the propagation of light in
the electrons' vicinity is sufficient to explain the electrons'
(gravitational) attraction in terms of non-resistant motion which is not
caused by a force. In such a way, as we have seen above, the case of the
classical electron may provide an insight into the question of why no force
is involved in the gravitational attraction of bodies as described by
general relativity.

\section{Looking for the origin of inertia in QFT}

As discussed in the Introduction the same mechanism that accounts for
inertia and mass of the classical electron - the interaction of the electron
charge with its distorted field - appears to lead to contributions from all
interactions to the inertia and mass in the framework of QFT.

Consider a free particle, whose constituents are subjected to
electromagnetic, weak, and strong interactions. The recoils the particle
suffers due to the emission and absorption of the virtual quanta of the
electromagnetic, weak, and strong "fields" of its own constituents cancel
out precisely. However, it has been overlooked that the recoils from the
virtual quanta absorbed by a non-inertial particle do not cancel - owing to
the general relativistic directional dependency of the frequencies of the
incoming virtual quanta (constituting a distortion of the corresponding
"fields") the magnitudes of their momenta and therefore the recoils acting
on the particle are also direction dependent. That imbalance in the recoils
gives rise to a self-force which is electromagnetic, weak, and strong in
origin and which acts on the non-inertial particle. As this self-force
arises only when the particle is prevented from following a geodesic path it
is clearly inertial.

At the conceptual level it appears certain that QFT does say something
important about the origin of inertia and mass - if one believes in the
Standard Model they should agree that the recoils from all incoming (blue or
red shifted) virtual quanta absorbed by the electric, weak, and strong
non-inertial charges will not cancel out and will give rise to an inertial
force. However, a conceptual analysis does not constitute a real result in
physics.

What is not only difficult, but almost hopeless is how one can calculate the
self-force originating from the unbalanced recoils from the virtual quanta
that are absorbed by a non-inertial particle. There are too many unknowns.
Virtual quanta are off mass shell particles which means that their energy,
momentum and mass do not obey the relativistic equation:

\[
E^{2}=p^{2}c^{2}+\left( mc^{2}\right) ^{2}.
\]%
For example, unlike the momentum of a normal photon which is given by $p=E/c$%
, the momentum of a virtual photon is not equal to $E/c$. Also unknown are
the energy of a charge (electromagnetic, weak, or strong) in terms of the
energies of the virtual quanta that constitute the "field" of the charge,
the lifetimes and energies of individual virtual quanta, and the absorption
time during which a virtual quant is absorbed by the charge.

An attempt to overcome those difficulties is to carry out a semiclassical
calculation of the self-force in the case of electromagnetic interaction in
QED; similar calculations can be done for the other interactions. For this
purpose the following assumptions will be made. It appears natural to define
the energy of the electric field of a charge in QED as the sum of the
energies of all virtual photons constituting the "field" at a given moment.
Such a definition avoids the issue of the dimension of the charge. In order
to eliminate some of the unknowns mentioned above it appears that an
equivalent definition of the charge's "field" energy is also possible in
terms of the total energy of the number of virtual photons absorbed by the
charge during some characteristic time $\delta t$ ($\delta t$ can be
regarded as the time for which the charge renews its "field"). The lifetimes
of all virtual photons can be expressed in terms of $\delta t$ as $\alpha
\delta t$, where $\alpha $ is a real number. The distances travelled by the
virtual photons during their lifetimes will be then $\alpha r=\alpha c\delta
t$, where $r=c\delta t$, is obviously the distance travelled by a virtual
photon for the characteristic time $\delta t$. The time for which a virtual
photon is absorbed by a charge is assumed to be $\delta \tau $. The momentum
of a virtual photon can be considered to be $p^{a}=\beta \left(
E^{a}/c\right) $, where $\beta $ is also a real number and $E^{a}$ is the
blue/red shifted energy of the virtual photon being absorbed by an
accelerating charge; $E^{a}$ is determined in the accelerating reference
frame $N^{a}$ in which the charge is at rest.

In $N^{a}$ the frequency of a virtual photon coming from a given directions
toward the charge (as seen by the charge) can be written in the vector form

\begin{equation}
f^{a}=f\left( 1-\frac{\mathbf{a\cdot r}}{c^{2}}\right) .  \label{f_a}
\end{equation}%
Here $f$ is the frequency measured at $\mathbf{r=}$ $0$ and $\mathbf{r=}$ $%
\mathbf{n}r$, where $\mathbf{n}$ is a unit vector pointing towards the
charge and determining the direction of the incoming virtual photon.

By the uncertainty principle the energy of a virtual photon of lifetime $%
\delta t$ in $N^{a}$ is $\Delta E^{a}\sim $%
h{\hskip-.2em}\llap{\protect\rule[1.1ex]{.325em}{.1ex}}{\hskip.2em}%
$/\delta t$. A virtual photon of lifetime $\alpha \delta t$ will have energy
$\Delta E_{\alpha }^{a}\sim $%
h{\hskip-.2em}\llap{\protect\rule[1.1ex]{.325em}{.1ex}}{\hskip.2em}%
$/\alpha \delta t=\Delta E^{a}/\alpha $. By (\ref{f_a}) the energy of this
virtual photon can be written as

\[
\Delta E_{\alpha }^{a}=\frac{\Delta E^{a}}{\alpha }=\frac{hf^{a}}{\alpha }=%
\frac{hf}{\alpha }\left( 1-\frac{\mathbf{a\cdot r}}{c^{2}}\alpha \right) =%
\frac{\Delta E}{\alpha }\left( 1-\frac{\mathbf{a\cdot r}}{c^{2}}\alpha
\right) ,
\]%
where $\Delta E/\alpha =hf/\alpha $ \ is the energy of that virtual photon
determined at $\mathbf{r=}$ $0$. The momentum of the virtual photon will be
then

\[
\Delta p_{\alpha \beta }^{a}=\beta \frac{\Delta E_{\alpha }^{a}}{c}=\beta
\frac{\Delta E^{a}}{c\alpha }=\frac{\Delta E\beta }{c\alpha }\left( 1-\frac{%
\mathbf{a\cdot r}}{c^{2}}\alpha \right) .
\]

Assume that the number of virtual photons coming from a direction $\mathbf{n}
$ within the solid angle $d\Omega $ which are absorbed during the
characteristic time $\delta t$ is $x$. The momentum of all virtual photons $%
x $ is then

\[
\sum\limits_{i=1}^{x}\Delta p_{\alpha _{i}\beta _{i}}^{a}\mathbf{n~}d\Omega
=\sum\limits_{i=1}^{x}\frac{\Delta E^{a}\beta _{i}}{c\alpha _{i}}\mathbf{n~}%
d\Omega =\sum\limits_{i=1}^{x}\frac{\Delta E\beta _{i}}{c\alpha _{i}}\left(
1-\frac{\mathbf{a\cdot r}}{c^{2}}\alpha _{i}\right) \mathbf{n~}d\Omega .
\]

The force produced by the recoils from all $x$ virtual photons is

\[
d\mathbf{F}_{self}^{a}=\sum\limits_{i=1}^{x}\frac{\Delta p_{\alpha _{i}\beta
_{i}}^{a}}{\delta \tau }\mathbf{n~}d\Omega .
\]

Then the self-force that acts on the charge and results from the unbalanced
recoils of all virtual photons absorbed during the characteristic time $%
\delta t$ and coming from all directions towards the charge is

\begin{eqnarray}
\mathbf{F}_{self}^{a} &=&\int \sum\limits_{i=1}^{x}\frac{\Delta p_{\alpha
_{i}\beta _{i}}^{a}}{\delta \tau }\mathbf{n\ }d\Omega =\int
\sum\limits_{i=1}^{x}\frac{\Delta E\beta _{i}}{c\delta \tau \alpha _{i}}%
\left( 1-\frac{\mathbf{a\cdot r}}{c^{2}}\alpha _{i}\right) \mathbf{n\ }%
d\Omega  \nonumber \\
&=&\int \sum\limits_{i=1}^{x}\frac{\Delta E\beta _{i}}{c\delta \tau \alpha
_{i}}\mathbf{n\ }d\Omega -\int \sum\limits_{i=1}^{x}\frac{\Delta E\beta _{i}%
}{c^{3}\delta \tau }\left( \mathbf{a\cdot r}\right) \mathbf{n\ }d\Omega .
\label{vp22}
\end{eqnarray}%
Due to symmetry, the first integral in (\ref{vp22}) is zero. Noticing that $%
\mathbf{r=}$ $\mathbf{n}r$ and $r=c\delta t$, for the self-force we can write

\begin{equation}
\mathbf{F}_{self}^{a}=-\sum\limits_{i=1}^{x}\frac{\Delta E\beta _{i}\delta t%
}{c^{2}\delta \tau }\int \left( \mathbf{a\cdot n}\right) \mathbf{n\ }d\Omega
.  \label{vp3}
\end{equation}%
The integral in (\ref{vp3}) is similar to the one calculated in the Appendix:

\[
\int \left( \mathbf{a\cdot n}\right) \mathbf{n\ }d\Omega =\frac{4\pi }{3}%
\mathbf{a.}
\]%
By substituting this result in (\ref{vp3}) and taking into account that

\[
U=4\pi \sum\limits_{i=1}^{x}\Delta E\beta _{i}
\]%
is the total energy of all virtual photons (coming from all directions of
the solid angle $4\pi $) absorbed during the time $\delta t$, we obtain for
the self-force

\[
\mathbf{F}_{self}^{a}=-\frac{1}{3}\frac{\delta t}{\delta \tau }\frac{U}{c^{2}%
}\mathbf{a}.
\]%
The energy $U$ represents the energy of the "field" of the charge. Therefore
the quantity $U/c^{2}$ is the mass corresponding to the energy of the
electric "field" of the charge and can be regarded as its electromagnetic
mass $m^{a}=U/c^{2}$. If the characteristic time $\delta t$ is three times
greater than the absorption time $\delta \tau $, the expression for the
self-force acting on the accelerating charge will have the exact form of the
inertial force

\begin{equation}
\mathbf{F}_{self}^{a}=-m^{a}\mathbf{a.}  \label{vp4}
\end{equation}%
Despite the fact that the calculations are semiclassical everything in the
self-force (\ref{vp4}) shows that it can be regarded as a fraction of the
inertial force to which an accelerating charge is subjected - it is
proportional to the acceleration with the correct sign and the coefficient
of proportionality has the dimension of mass. The mass $m^{a}$ in (\ref{vp4}%
) is inertial since it is a measure of the resistance the charge offers to
its acceleration and is electromagnetic in origin.

Consider now a charge at rest in a gravitational field of strength $\mathbf{g%
}$. The frequency of a virtual photon absorbed by the charge will be shifted:

\begin{equation}
f^{g}=f\left( 1+\frac{\mathbf{g\cdot r}}{c^{2}}\right) .  \label{f-g}
\end{equation}%
If the virtual photon is approaching the charge from "above" (moving towards
the mass producing the gravitational field), the photon will be blue
shifted; if it is approaching the charge from "below" (receding from the
mass), it will be red shifted.

Like in the case of an accelerating charge, the unbalanced recoils of the
virtual photons that are absorbed during the time $\delta t$ by a charge at
rest in a gravitational field give rise to a self-force, which has precisely
the form of the gravitational force (if $\delta t=3\delta \tau $)

\begin{equation}
\mathbf{F}_{self}^{g}=m^{g}\mathbf{g.}  \label{vp5}
\end{equation}%
Here also $m^{g}=U/c^{2}$, where

\[
U=4\pi \sum\limits_{i=1}^{x}\Delta E\beta _{i}
\]%
is the total energy of all virtual photons (coming from all directions of
the solid angle $4\pi $) and absorbed by the non-inertial charge during the
time $\delta t$.

Everything in the self-force (\ref{vp5}) indicates that it can be regarded
as a fraction of the gravitational force acting on a charge supported in a
gravitational field - it is proportional to the gravitational acceleration
with the correct sign and the coefficient of proportionality has the
dimension of mass. It is clearly seen that the self-force (\ref{vp5}) is
inertial - it arises when the charge is prevented from following a geodesic
worldline (i.e. prevented from falling in the gravitational field); only in
this case the charge will see the incoming virtual photons as blue or red
shifted, i.e. with frequencies given by (\ref{f-g}). The mass $m^{g}$ in (%
\ref{vp5}) is traditionally regarded as passive gravitational mass but since
it is a measure of the resistance the charge is offering when deviated from
its geodesic path, $m^{g}$ is obviously inertial and in this case
electromagnetic in origin. That is why the masses $m^{a}$ and $m^{g}$
coincide - they are simply the same thing: (i) a measure of the resistance a
non-inertial charge is offering when prevented from following a geodesic
path, and (ii) the mass that correspond to the charge's "field".

When the charge is falling in the gravitational field the frequencies of the
incoming virtual photons will not be shifted as seen by the charge. The
recoils from the absorbed photons will cancel out and the charge will move
in a non-resistant manner - its worldline will be geodesic.

The equations (\ref{vp4}) and (\ref{vp5}) have the form of Newton's second
law. On the one hand, it is clear that in QED the behaviour of a charge is
not governed by the deterministic Newton's second law. On the other hand,
however, a non-inertial charge in QED is also subjected to an inertial or
gravitational force which should have the form of Newton's second law.

The outlined QED mechanism for the inertia and mass of a charge appears to
resemble the zero-point field (ZPF) approach to inertia \cite{haisch}, \cite%
{rueda}. In both approaches inertia originates from the interaction of an
accelerating charge with virtual photons. In the ZPF approach the resistance
an accelerated charge offers to its acceleration originates from its
interaction with the virtual photons of the ZPF fluctuations of the
electromagnetic quantum vacuum, whereas in the proposed mechanism it is the
unbalanced recoils from the virtual photons of the charge's own
electromagnetic field that give rise to the charge's inertia. The latter
mechanism provides a common explanation of the origin of the inertial and
the passive gravitational mass. The ZPF approach to inertia, however,
appears to encounter some difficulties with the explanation of the passive
gravitational mass. In a recent attempt \cite{hr} to overcome those
difficulties it has been assumed that "the electromagnetic quantum vacuum is
effectively accelerating (falling)" past a charge held fixed in a
gravitational field. That assumption, however, needs a thorough examination
since a falling vacuum implies that light is also falling (accelerating) in
a gravitational field which is not the case in general relativity; the
velocity of a light signal propagating (falling) towards a massive body is
decreasing \cite[p. 197]{ohanian}, \cite{petkov2}.

The weak and the strong interactions will make similar contributions to the
inertia and mass of a particle whose constituents are involved in weak and
strong interactions. Consider, for example, one of the quarks in a proton.
The quark participates in electromagnetic, weak, and strong interactions
since it has electric, weak (because its flavor can be altered only by
charged weak interactions), and color (strong) charges. If for the sake of
the argument we assume that the quark is not interacting with the other
quarks in the proton, its mass will be simply a sum of positive
electromagnetic, weak, and strong contributions. When the interactions
between the quarks is turned on the picture becomes complicated - as
discussed in Section 5 depending on the interactions (whether they are
repulsion or attraction) and on the angle between the acceleration ($\mathbf{%
a}$ or $\mathbf{g}$) and the line on interaction between the quarks, the
contribution of different interactions to the mass of the proton may be
positive or negative.

Let us summarize the behavior of a particle whose constituents are involved
in electromagnetic, weak, and strong interactions. If the recoils from all
virtual quanta that are absorbed by the particle's constituents cancel out,
the particle will move in a non-resistant way (by inertia) and its worldline
will be geodesic. If the particle is prevented from following a geodesic
path the balance in the recoins is disturbed and every constituent of the
particle and therefore the particle itself is subjected to a self-force
which is inertial in origin since it resists the deviation of the particle
from its geodesic path.

The particle will fall in a gravitational field of strength $\mathbf{g}$
with an acceleration $\mathbf{g}$ \emph{by itself} in order to prevent a
disturbance of the balance in the recoils (from virtual photons, $W$ and $Z$%
\ particles, and gluons) suffered by its constituents. Therefore all kinds
of particles should fall with the \emph{same} acceleration in a
gravitational field and should offer no resistance to their fall. It is
clear from here that if all the mass is composed of electromagnetic, weak
and strong contributions, the mechanism that is responsible for the
same-acceleration non-resistant fall of a particle simultaneously \emph{%
explains} two facts: (i) that in general relativity the motion of a body
falling towards a gravitating center is regarded as inertial (non-resistant)
and is represented by a geodesic worldline, and (ii) the experimental fact
that all objects fall in a gravitational field with the \emph{same}
acceleration.

One may be left with the impression that QFT may lead to a solution of the
mystery of inertia. Unfortunately, it is not that certain. What we have done
in this section is to explain inertia in terms of unbalanced recoils from
absorbed virtual quanta. However, the very existence of recoils seems to
imply that they are caused by the \textit{inertia} of the virtual quanta.
For instance, hardly anyone questions the inertia of the normal photon \cite%
{einstein1906}, \cite{livesey}. But even in this case, QFT\ may account for
the macroscopic manifestations of inertia.

\section{Conclusions}

The paper is concerned with two main issues. It revisits the classical
electromagnetic mass theory and further develops it by taking into account a
corollary of general relativity - that the propagation of light in
non-inertial reference frames is anisotropic. The arguments against
regarding the entire inertia and mass of the classical electron as
electromagnetic in origin have been answered.

The second issue addressed in the paper is an overlooked effect in QFT that
the accepted mechanism of interactions through exchange of virtual quanta
leads to acceleration-dependent self-interaction effects which may account
for the origin of inertia and mass of macroscopic objects.

The main results of the paper are:

1. The study of an accelerating electron and an electron supported in a
gravitational field in terms of the classical electromagnetic mass theory
yields three results: (i) There appears to exist a connection between the
shape of the electric field of a charge and the shape of its worldline. The
worldline of a charge whose field is distorted (due to the charge's
acceleration or its being at rest in a gravitational field) is not geodesic
and the charge resists the deformation of its field (and worldline) - the
charge is subjected to a self-force on account of its distorted field which
is electromagnetic in origin. If the field of a charge is the Coulomb field
its worldline is geodesic and the charge offers no resistance to its motion
while moving with constant velocity in flat spacetime or falling in a
gravitational field. (ii) The self-force acting on an accelerating electron
or on an electron supported in a gravitational field is \textit{inertial}
and is electromagnetic in origin. Therefore, the equivalence of the inertial
and gravitational forces and masses (in the case of the classical electron)
is naturally explained. (iii) Newton's second law is derived on the basis of
Maxwell's equations and the classical model of the electron.

2. It has been shown that the average velocity of light propagating in
non-inertial reference frames is anisotropic. This anisotropy (i) makes it
possible to calculate the electric field and self-force of a charge \emph{%
directly} in a non-inertial reference frame without the need to transform
the field from an inertial frame, and (ii) sheds some light on what might be
the underlying reason of the equivalence principle.

3. The anisotropic velocity of light in $N^{a}$ or $N^{g}$ gives rise to an
anisotropic volume element $dV^{a}$ or $dV^{g}$ of a charge determined in $%
N^{a}$ or $N^{g}$, respectively. It is the use of this anisotropic volume
element that yields the correct expressions of the potential and the
electric field of a charge in non-inertial frames - taking the anisotropic
volume element into account simultaneously solves two problems: removes (i)
the $1/2$ factor in the potential of a charge supported in a gravitational
field (calculated by Fermi in 1921), and (ii) the $4/3$ factor in the
self-force.

4. It has been shown that in QFT a mechanism, similar to the mechanism
giving rise to inertia and mass of the classical electron, also leads to
contributions from the electromagnetic, weak, and strong interactions to
inertia and mass.

\section{Appendix - Calculation of the Self-Force}

The self-force
\[
\mathbf{F}_{self}^{a}=\frac{1}{4\pi \epsilon _{o}}\int \int \left( \frac{%
\mathbf{n}}{r^{2}}+\frac{\mathbf{a}\cdot \mathbf{n}}{c^{2}r}\mathbf{n}-\frac{%
1}{c^{2}r}\mathbf{a}\right) \left( 1+\frac{\mathbf{a}\cdot \mathbf{r}}{2c^{2}%
}\right) \rho ^{2}dVdV_{1}{.}
\]
can be written (to within terms proportional to $c^{-2}$) as
\begin{equation}
\mathbf{F}_{self}^{a}=\frac{1}{4\pi \epsilon _{o}}\int \int \left( \frac{%
\mathbf{n}}{r^{2}}+\frac{3}{2}\frac{\mathbf{a}\cdot \mathbf{n}}{c^{2}r}%
\mathbf{n}-\frac{1}{c^{2}r}\mathbf{a}\right) \rho ^{2}dVdV_{1}{.}
\label{1force_a}
\end{equation}

We have reached this result assuming that the charge element $de^{a}$ acts
upon the charge element $de_{1}^{a}$. In this case the vector $\mathbf{r}$
begins at $de^{a}$ and ends at $de_{1}^{a}$, i.e. $\mathbf{n}$ points from $%
de^{a}$ to $de_{1}^{a}$. If we assumed that $de_{1}^{a}$ acts upon $de^{a}$
the result should be the same. As interchanging the two charge elements
reverses the direction of $\mathbf{n}$ the self-force in this case will be

\begin{equation}
\mathbf{F}_{self}^{a}=\frac{1}{4\pi \epsilon _{o}}\int \int \left( -\frac{%
\mathbf{n}}{r^{2}}+\frac{3}{2}\frac{\mathbf{a}\cdot \mathbf{n}}{c^{2}r}%
\mathbf{n}-\frac{1}{c^{2}r}\mathbf{a}\right) \rho ^{2}dVdV_{1}{.}
\label{force_a1}
\end{equation}%
Adding equations (\ref{force_a1}) and (\ref{1force_a}) and dividing the
result by 2 we get
\begin{equation}
\mathbf{F}_{self}^{a}=\frac{1}{4\pi \epsilon _{o}}\int \int \left( \frac{3}{2%
}\frac{\mathbf{a}\cdot \mathbf{n}}{c^{2}r}\mathbf{n}-\frac{1}{c^{2}r}\mathbf{%
a}\right) \rho ^{2}dVdV_{1}.  \label{force}
\end{equation}%
In order to do the integral (\ref{force}) let us consider the integral \cite%
{panofski}
\begin{equation}
\mathbf{I}=\int \int \left( \frac{\mathbf{a}\cdot \mathbf{n}}{r}\mathbf{n}%
\right) dVdV_{1}.  \label{i1}
\end{equation}%
We can put $\mathbf{n=n}_{\parallel }\mathbf{+n}_{\perp }$, where $\mathbf{n}%
_{\parallel }$ is parallel to $\mathbf{a}$ and $\mathbf{n}_{\perp }$ is
perpendicular to $\mathbf{a}$. Then
\begin{eqnarray*}
\left( \mathbf{a}\cdot \mathbf{n}\right) \mathbf{n} &=&\mathbf{a}\left(
\mathbf{n}_{\parallel }\mathbf{+n}_{\perp }\right) \left( \mathbf{n}%
_{\parallel }\mathbf{+n}_{\perp }\right) \\
&=&\left( \mathbf{a}\cdot \mathbf{n}_{\parallel }\mathbf{+a}\cdot \mathbf{n}%
_{\perp }\right) \left( \mathbf{n}_{\parallel }\mathbf{+n}_{\perp }\right) \\
&=&\left( \mathbf{a}\cdot \mathbf{n}_{\parallel }\right) \mathbf{n}%
_{\parallel }+\left( \mathbf{a}\cdot \mathbf{n}_{\parallel }\right) \mathbf{n%
}_{\perp }+\left( \mathbf{a}\cdot \mathbf{n}_{\perp }\right) \mathbf{n}%
_{\parallel }+\left( \mathbf{a}\cdot \mathbf{n}_{\perp }\right) \mathbf{n}%
_{\perp } \\
&=&\left( \mathbf{a}\cdot \mathbf{n}_{\parallel }\right) \mathbf{n}%
_{\parallel }+\left( \mathbf{a}\cdot \mathbf{n}_{\parallel }\right) \mathbf{n%
}_{\perp }
\end{eqnarray*}%
since $\left( \mathbf{a}\cdot \mathbf{n}_{\perp }\right) =0$. Substituting
this result in (\ref{i1}) yields
\begin{equation}
\mathbf{I}=\int \int \left( \frac{\mathbf{a}\cdot \mathbf{n}_{\parallel }}{r}%
\mathbf{n}_{\parallel }\right) dVdV_{1}+\int \int \left( \frac{\mathbf{a}%
\cdot \mathbf{n}_{\parallel }}{r}\mathbf{n}_{\perp }\right) dVdV_{1}.
\label{i2}
\end{equation}%
To facilitate the calculations further let us assume that $\mathbf{r}$ is
rotated 180${{}^{\circ }}$ about an axis parallel to $\mathbf{a}$ running
through the centre of the spherical charge distribution of the electron.
Then the vector $\mathbf{n=n}_{\parallel }\mathbf{+n}_{\perp }$ becomes $%
\mathbf{n}_{\parallel }\mathbf{-n}_{\perp }$. This means that in the second
integral in (\ref{i2}) for every elementary contribution

\[
\left( \frac{\mathbf{a}\cdot \mathbf{n}_{\parallel }}{r}\mathbf{n}_{\perp
}\right) dVdV_{1}
\]
there is also an equal and opposite contribution

\[
-\left( \frac{\mathbf{a}\cdot \mathbf{n}_{\parallel }}{r}\mathbf{n}_{\perp
}\right) dVdV_{1}
\]
which shows that the second integral in (\ref{i2}) is zero and we can write
\begin{equation}
\mathbf{I}=\int \int \left( \frac{\mathbf{a}\cdot \mathbf{n}_{\parallel }}{r}%
\mathbf{n}_{\parallel }\right) dVdV_{1}.  \label{i3}
\end{equation}

The integral $\mathbf{I}$ is now a function only of $\mathbf{n}_{\parallel }$%
. In order to return to the general case of $\mathbf{n}$ (and not restrict
ourselves to using $\mathbf{n}_{\parallel }$) we will express the integral
in (\ref{i3}) in terms of $\mathbf{n}$ and a unit vector $\mathbf{u}$ in the
direction of $\mathbf{a}$. Since $\mathbf{n}_{\parallel }$ is parallel to $%
\mathbf{a}$, we have $\mathbf{a}\cdot \mathbf{n}_{\parallel }=an_{\parallel
} $. Then we can write
\begin{eqnarray*}
\left( an_{\parallel }\right) \mathbf{n}_{\parallel } &=&\mathbf{a}\left(
n_{\parallel }\right) ^{2}=\mathbf{a}\left[ 1^{2}\left( n_{\parallel
}\right) ^{2}\right] = \\
&=&\mathbf{a}\left( 1n_{\parallel }\right) ^{2}=\mathbf{a}\left(
un_{\parallel }\right) ^{2}=\mathbf{a}\left( un\cos \theta \right) ^{2} \\
&=&\mathbf{a}\left( \mathbf{u\cdot n}\right) ^{2}
\end{eqnarray*}
where $\theta $ is the angle the vector $\mathbf{n}$ forms with the vector
of the acceleration $\mathbf{a}$. Now we can write the integral (\ref{i3})
in the form
\begin{equation}
\mathbf{I}=\mathbf{a}\int \int \frac{\left( \mathbf{u}\cdot \mathbf{n}%
\right) ^{2}}{r}dVdV_{1}.  \label{i4}
\end{equation}

Following Abraham \cite{abraham} and Lorentz \cite{lorentz} we have assumed
a spherically symmetric distribution of the electron charge. This shows that
as all directions in space are indistinguishable the integral in (\ref{i4})
should be independent of the direction of the unit vector $\mathbf{u}$. In
such a way the average of this integral over all possible directions of $%
\mathbf{u}$ should be equal to the integral itself:
\begin{eqnarray}
\int \int \frac{\left( \mathbf{u}\cdot \mathbf{n}\right) ^{2}}{r}dVdV_{1} &=&%
\frac{1}{4\pi }\int d\Omega \int \int \frac{\left( \mathbf{u}\cdot \mathbf{n}%
\right) ^{2}}{r}dVdV_{1}  \label{i5} \\
&=&\frac{1}{4\pi }\int \int \frac{dVdV_{1}}{r}\int \left( \mathbf{u}\cdot
\mathbf{n}\right) ^{2}d\Omega  \nonumber
\end{eqnarray}%
where $d\Omega $ is an element of the solid angle within which a given unit
vector $\mathbf{u}$ lies. To do this integral we choose a polar coordinate
system with the polar axis along $\mathbf{n}$. Then $\mathbf{u}\cdot \mathbf{%
n=}\cos \theta $ and $d\Omega =\sin \theta d\theta d\varphi $ and
\begin{eqnarray*}
\frac{1}{4\pi }\int \left( \mathbf{u}\cdot \mathbf{n}\right) ^{2}d\Omega &=&%
\frac{1}{4\pi }\int_{0}^{\pi }\cos ^{2}\theta \sin \theta d\theta
\int_{0}^{2\pi }d\varphi \\
&=&\frac{1}{2}\int_{0}^{\pi }\cos ^{2}\theta \sin \theta d\theta \\
&=&\frac{1}{2}\left( -\frac{1}{3}\cos ^{3}\theta \mid _{0}^{\pi }\right) \\
&=&\frac{1}{2}\left[ -\frac{1}{3}\left( -1-1\right) \right] \\
&=&\frac{1}{3}{.}
\end{eqnarray*}%
Substituting this result in (\ref{i5}) yields

\[
\int \int \frac{\left( \mathbf{u}\cdot \mathbf{n}\right) ^{2}}{r}dVdV_{1}=%
\frac{1}{3}\int \int \frac{dVdV_{1}}{r}.
\]%
Thus for the integral (\ref{i4}) we have
\begin{equation}
\mathbf{I}=\mathbf{a}\int \int \frac{\left( \mathbf{u}\cdot \mathbf{n}%
\right) ^{2}}{r}dVdV_{1}=\frac{\mathbf{a}}{3}\int \int \frac{dVdV_{1}}{r}.
\label{i6}
\end{equation}%
By substituting (\ref{i6}) in (\ref{force}) we obtain
\begin{eqnarray*}
\mathbf{F}_{self}^{a} &=&\frac{1}{4\pi \epsilon _{o}}\int \int \left( \frac{3%
}{2}\frac{\mathbf{a}}{3c^{2}r}-\frac{\mathbf{a}}{c^{2}r}\right) \rho
^{2}dVdV_{1} \\
&=&-\frac{\mathbf{a}}{8\pi \epsilon _{o}c^{2}}\int \int \frac{\rho ^{2}}{r}%
dVdV_{1}
\end{eqnarray*}%
and finally the expression for the self-force becomes
\[
\mathbf{F}_{self}^{a}=-\frac{U}{c^{2}}\mathbf{a}{.}
\]


\begin{thebibliography}{99}
\bibitem{4/3factor} In fact, the obtained relationship between mass and
energy contained a factor of $4/3$ which later was accounted for; see \cite[%
p. 28-4]{feynman}.
\bibitem{charge} The fact that the volume of the charge depends on the
observation point at which the potential is determined does not mean that
there is a violation of the invariance of the electric charge. For a
specific discussion of this point see \cite[p. 342]{panofski}; for a
derivation of the Li\'{e}nard-Wiechert volume in the calculation of the Li%
\'{e}nard-Wiechert potentials see \cite[p. 419]{griffiths}.
\end{thebibliography}
\end{document}